\documentclass[english]{article}

\pdfoutput=1
\usepackage{lmodern}
\usepackage{booktabs}
\usepackage[T1]{fontenc}
\usepackage{subfloat}
\usepackage[latin9]{inputenc}
\usepackage{babel}

\makeatletter

\usepackage{mathrsfs}   
\usepackage{slashed}     
\usepackage{url}
\usepackage{graphicx}
\usepackage{color} 
\usepackage{bm}
\usepackage[numbers,sort&compress]{natbib}
\usepackage{float}
\usepackage{multirow}
\usepackage{amsmath}
\usepackage{amssymb}
\usepackage{breqn}
\usepackage{subcaption}
\usepackage{caption}
\usepackage{mciteplus}
\usepackage{authblk}
\usepackage{diagbox}
\usepackage{stackengine}
\usepackage[normalem]{ulem}

\usepackage{tikz}

\usetikzlibrary{decorations.markings}
\usetikzlibrary{decorations.pathmorphing}

\tikzset{
    vector/.style={decorate, decoration={snake}, draw},
    fermion/.style={draw=black, postaction={decorate},
        decoration={markings,mark=at position .55 with {\arrow[scale=1.,>=stealth]{>}}}},
    fermionbar/.style={draw=black, postaction={decorate},
        decoration={markings,mark=at position .58 with {\arrow[scale=1.,>=stealth]{<}}}},
    fermionline/.style={draw=black},
    gluon/.style={decorate, draw=black,
        decoration={coil,amplitude=4pt, segment length=5pt}},
    scalar/.style={dashed, draw=black, postaction={decorate},
        decoration={markings,mark=at position .55 with {\arrow[scale=1.,>=stealth]{>}}}},
    scalarbar/.style={dashed, draw=black, postaction={decorate},
        decoration={markings,mark=at position .58 with {\arrow[scale=1.,>=stealth]{<}}}},
    scalarline/.style={dashed,draw=black},
    fmassin/.style={draw=black, postaction={decorate},
    decoration={markings, mark=at position .75 with {\arrow[scale=1.,>=stealth]{>}}, mark=at position .30 with {\arrow[scale=1.,>=stealth]{<}}}},
}

\usepackage[colorlinks=true,linkcolor=redLinks,citecolor=greenLinks,urlcolor=redLinks, pdfborder={0 0 1}]{hyperref}
\definecolor{greenLinks}{rgb}{0, 0.6, 0} 
\definecolor{blueLinks}{rgb}{0, 0, 0.6}
\definecolor{redLinks}{rgb}{0.6, 0, 0}
\definecolor{eprintLinks}{rgb}{0.4, 0.4, 0.4}
\definecolor{journalLinks}{rgb}{0.6, 0, 0}

\begin{document}

\title{\bf $J/\psi$ production at NLO with a scale-dependent color-evaporation model}
\author[1]{B. Guiot\thanks{benjamin.guiot@usm.cl}}
\author[1]{A. Radic}
\author[1]{I. Schmidt}
\author[2]{K. Werner}
\affil[1]{Departamento de F\'isica, Universidad T\'ecnica Federico Santa Mar\'ia; Casilla 110-V, Valparaiso, Chile}
\affil[2]{SUBATECH, University of Nantes - IN2P3/CNRS - IMT Atlantique, Nantes, France}

\renewcommand\Authands{ and }
\date{}

\maketitle
\begin{abstract}
Nearly ten years ago, Kang, Ma, Qiu, and Sterman derived an evolution equation for a $Q\bar{Q}$ pair fragmenting into a  quarkonium. In this study we explore the consequence of this evolution for the color-evaporation model, focusing on $J/\psi$ transverse-momentum ($p_t$) distributions in proton-proton collisions. We show that, as expected, it softens the spectrum obtained by fixed-order calculations. While next-to-leading-order calculations strongly overestimate data at large $p_t$, ours, including the (approximate) $Q\bar{Q}$ evolution and next-to-leading-order cross sections computed with Madgraph, are in good agreement with experiments. Since our study with the color-evaporation model shows a significant effect of the $Q\bar{Q}$ evolution at large $p_t$, a new determination of long-distance-matrix elements of nonrelativistic QCD could be necessary. To describe data at small and intermediate $p_t$, we use the $k_t$-factorization approach, and we argue that quarkonia data could help constrain this formalism.
\end{abstract}
\newpage

\tableofcontents

\section{Introduction}

Quarkonia production is important in Quantum Chromodynamics (QCD) because it provides insights into the fundamental dynamics of QCD, the structure of the QCD vacuum, the effects of heavy quark masses, the properties of the quark-gluon plasma, and allows for experimental tests of theoretical predictions. Models for quarkonia production, generally based on a factorization hypothesis, differ by their treatment of hadronization. For instance, the nonrelativistic QCD (NRQCD) \cite{Bodwin1995} expression for the quarkonium cross section reads
\begin{equation}
    \sigma^H=\sum_{[Q\bar{Q}(n)]}\hat{\sigma}_{[Q\bar{Q}(n)]}(\Lambda)\langle 0| \mathcal{O}_{[Q\bar{Q}(n)]}^H(\Lambda)|0\rangle, \label{nrqcd}
\end{equation}
where $\hat{\sigma}_{[Q\bar{Q}(n)]}$ are the process-dependent short-distance coefficients to produce a heavy-quark pair in a quantum state $n$ (color, spin, and orbital angular momentum states). The second factor in the right side of Eq.~(\ref{nrqcd}) is the long-distance-matrix element (LDME) describing the non-perturbative hadronization of the $Q\bar{Q}(n)$ pair into the quarkonium $H$. Both factors depend on the ultraviolet cutoff $\Lambda \sim \mathcal{O}(m_Q)$, with $m_Q$ the heavy-quark mass. 

On the opposite, the CEM uses the same weight for all $Q\bar{Q}$ states and has only one non-perturbative parameter, the hadronization factor $F_{H}$. It is then natural to think that the CEM is not fully consistent with QCD, as discussed in Ref. \cite{Bodwin2005}. However, its simplicity and reasonably good description of numerous observables make it a useful tool for the study of quarkonia production.\\

\begin{figure}[!h]
\centering
\subcaptionbox{\label{nloctr}}
{\includegraphics[width=8pc]{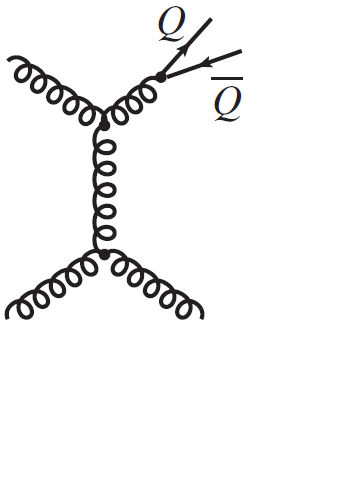}}
\subcaptionbox{\label{madfo}}
{\includegraphics[width=20pc]{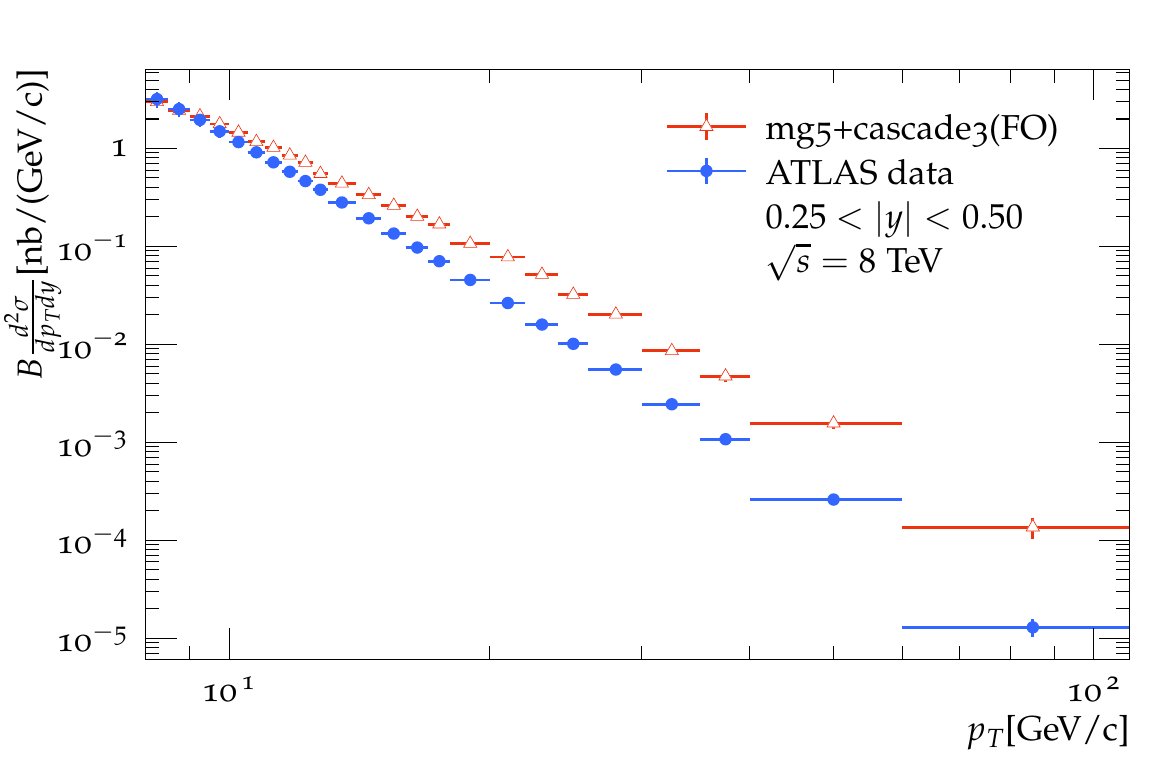}}
\caption{(a) Example of NLO contribution scaling as $1/p_t^4$. (b) Comparison of fixed-order calculations (red triangles) with ATLAS data (blue dots) for the transverse-momentum distribution of prompt $J/\psi$ \cite{Aad2016}. For this result, we used madgraph5\_aMC@NLO at NLO using a 3-flavor-number scheme and a charm mass $m_c=1.3$ GeV.}\label{nloexp}
\end{figure}

An issue of the CEM is the too-hard spectrum due to next-leading order (NLO) contributions scaling as $1/p_t^4$ \cite{Lansberg2020}, where $p_t$ is the $J/\psi$ transverse momentum. An example of such contribution is shown in Fig.~\ref{nloctr}, along with the result of fixed-order NLO  calculations obtained with madgraph5\_aMC@NLO (MG5) \cite{Alwall2014} in Fig.~\ref{madfo}.
In this study, we demonstrate that the evolution of the heavy-quark pair, presented in Sec.~\ref{secevol}, brings CEM calculations in agreement with data for the $p_t$-distribution of the $J/\psi$ particle. This evolution, related to the scale-dependence of fragmentation functions, is a central prediction of perturbative QCD.\\

Our first goal is not to support the CEM, but to underline the importance of the heavy-quark pair evolution. Of course, having the accuracy of this model improved is also of interest. The present study suggests that evolution could significantly affect the determination of NRQCD's LMDEs. We could reasonably hope that a new extraction of these non-perturbative functions, including the evolution, will alleviate the tensions between different LDME sets and improve the NRQCD description of experimental data.\\

To describe data on the whole $p_t$ range, we use a transverse-momentum dependent formalism (see Sec.~\ref{secForm}). Our second goal is to use quarkonium data to put constraints on transverse-momentum-dependent PDFs. Our main results are presented in Sec.~\ref{secRes}, and we discuss the universality of the non-perturbative parameter, $F_{J/\psi}$, in Sec.~\ref{secUni}. Finally, we show similar results obtained with the event generator EPOS4 in Sec.~\ref{secepos}. The common features of EPOS4 with our calculations are the timelike cascade, giving the (approximate) scale evolution of the $Q\Bar{Q}$ pair, and its hadronization based on the CEM. The similar results obtained with two different formalisms demonstrate the robustness of our main conclusion.

\section{Evolution of the heavy-quark pair \label{secevol}}
Two groups published a series of papers discussing the scale evolution of quarkonia fragmentation functions. One of them worked with perturbative QCD \cite{Kang2011,Kang2012,Kang2014,Lee2022}, while the other used the soft-collinear-effective theory \cite{Fleming2012,Fleming2013}. Contributions to quarkonia production are first separated into leading power, scaling as $1/p_t^4$, and next-to-leading power (NLP), scaling as $1/p_t^6$, and then factorized. It is the factorization of NLP contribution that involves the convolution of a short-distance partonic cross section for the production of a $Q\bar{Q}$ pair with the modified double-parton fragmentation function of a $Q\bar{Q}$ pair into a quarkonium $H$. This fragmentation function obeys the following evolution equation
\begin{multline}
    \frac{\partial}{\partial \ln \mu^2} D_{[Q \bar{Q}(\kappa)] \rightarrow H}\left(z, \mu^2\right)=\frac{\alpha_s(\mu)}{2 \pi} \sum_n \int_z^1 \frac{d z^{\prime}}{z^{\prime}} P_{[Q \bar{Q}(n)] \rightarrow[Q \bar{Q}(\kappa)]}\left(\frac{z}{z^{\prime}}\right) \\ \times D_{[Q \bar{Q}(n)] \rightarrow H}\left(z^{\prime}, \mu^2 \right), \label{evoleq}
\end{multline}
where $\kappa$ and $n$ denote the possible quantum numbers of the $Q\bar{Q}$ pair, and $ P_{[Q \bar{Q}(n)] \rightarrow[Q \bar{Q}(\kappa)]}$ are the modified evolution kernels, see \cite{Kang2014} for more details.\\

For the standard choice $\mu=p_t$, we see that at intermediate and large transverse momentum the color-octet (CO) and color-singlet (CS) states mix under Eq.~(\ref{evoleq}). The different models for quarkonia production should be applied at $\mu_0\sim m_Q$, and the CEM uses a $\delta(z-1)$ as $z$ distribution \cite{Fritzsch1977,Halzen1977}. Naively, we expect the evolution to reduce the differences between the different production mechanisms. Indeed, a $Q\bar{Q}$ pair in a CS state at $\mu_0$ could have been produced in a CO state at $\mu=p_t$. The mixing of quantum states induced by the evolution step $Q \bar{Q}(n) \rightarrow Q \bar{Q}(\kappa)$ implies a mixing of the LO behavior $1/p_t^a$ ($a>0$) associated to these states. The naive (LO) $p_t$ dependence of the produced $Q\bar{Q}(n)$ is further modified by real emissions, partly responsible for the evolution equation, and resulting in energy loss. It is in fact this effect which will bring the CEM calculations in agreement with data.\footnote{The mixing of states with different quantum numbers $\kappa$ does not matter, since the CEM uses the same LDME for all $\kappa$.}\\

In our study, we did not work with the solution of Eq.~(\ref{evoleq}), but used the PYTHIA6 \cite{Sjoestrand2006} timelike cascade to evolve the heavy-quark pair. A timelike cascade is an implementation of the DGLAP equation \cite{GribovLipatov:1972,AltarelliParisi:1977,Dokshitzer:1977} for timelike partons, and a central tool for all realistic event generators. An important simplification made by these algorithms is that timelike partons do not interact\footnote{An exception to this statement are the interferences leading to angular ordering \cite{Ellis1996}.\label{noteao}}. They simply split into two partons until they reach the lower cutoff of the cascade. By evolving the heavy quark and antiquark independently, we neglect several Feynman diagrams in Sec.~IV C. of Ref. \cite{Kang2014}. For instance, (virtual) diagrams with a gluonic line between the quark and antiquark in the amplitude (and nothing in the conjugate amplitude) are not included. Interference terms, where the gluon is emitted by the quark in the amplitude and by the antiquark in the conjugate amplitude are also not strictly included. 

However, virtual corrections are sometimes negative, cancellations can occur and the importance of the neglected Feynman diagrams is unclear. From phenomenological knowledge, one may argue that their net contribution is not significant. Indeed, the discussion for the $Q\bar{Q}$ pair is also true for any hard parton producing a jet; we neglect the Feynman diagrams associated with the interactions between timelike particles forming the jet. Since event generators and timelike cascades provide a good description of data, including exclusive observables, it seems to be a reasonable approximation. From this point of view, the heavy quark and antiquark have nothing special. In the real world, after a few collinear splittings, they will be (closely) surrounded by other partons and interact with them. But the heavy quarks are evolved as any other parton, by neglecting these interactions, and, in the end, will hadronize either into D mesons or charmonium.\\

Thanks to the simplicity of the CEM, the absence of evolution is easily observable, since, as demonstrated in Sec.~\ref{secRes}, it is responsible for the overestimation of data by NLO calculations. This is another reason why we choose to work with the CEM. In the case of NRQCD, taking into account the evolution would certainly require a new determination of the LDMEs. The consequences of this could be modest for inclusive observables, and more visible for exclusive observables.

\section{Heavy-quark pair production, evolution and hadronization \label{secForm}}
In order to describe $J/\psi$ production on the whole $p_t$ range, we use a transverse-momentum dependent formalism. A possible choice would be the TMD factorization, discussed for quarkonia production in proton-proton (pp) collisions using the soft-collinear and NRQCD effective field theories \cite{Echevarria2019}. Another possibility is the $k_t$ factorization approach \cite{Grlery,Catani1990,Collins1991,Catani1991,Lerysh}. This formalism has been extensively used for the study of $J/\psi$ production in pp collisions \cite{Baranov2002,Jung2011,Saleev2012,Baranov2012,Cheung2018,Cisek2018,Maciula2019,Baranov2019,Chernyshev2022}. In our case, we used a hybrid formalism, already used for Drell-Yan \cite{Martinez2020}, where the off-shell cross section is replaced by on-shell cross section. For proton-proton collisions, the differential cross section reads
\begin{multline}
 \frac{d\sigma(pp\to J/\psi + X)}{dx_1dx_2d^2p_t}(s,x_1,x_2,p_t)=\sum_{a,b}\int^{k_{t,\text{max}}^2\sim s}_0 d^2k_{1t}d^2k_{2t} F_{a/p}(x_1,k_{1t};\mu)\\
 \times F_{b/p}(x_2,k_{2t};\mu)d\hat{\sigma}_{ab\to Q\Bar{Q}}(x_1x_2s,q_t;\mu)\otimes d_{Q\bar{Q}}(\mu,\mu_0)\otimes D^{\text{CEM}}_{Q\bar{Q}\to J/\psi}(\mu_0), \label{ktfac}
\end{multline}
where $\bm{q}_t=\bm{p}_t-\bm{k}_{1t}-\bm{k}_{2t}$, with $\bm{k}_{1t}$ and $\bm{k}_{2t}$ the transverse momentum of the two incoming partons. The variables $x_{1,2}$ are the longitudinal momentum fractions carried by the partons, $\mu$ is the factorization scale, and $s$ is the usual Mandelstam variable for the proton-proton system. The functions $F_{i/p}$ are the unintegrated PDFs (UPDFs) for a parton $i$ inside a proton. The transverse-momentum dependence of the initial partons is managed with CASCADE3, and we used the parton-branching (PB) UPDFs, set 2 \cite{Martinez2019}, extracted from HERA data. They obey an evolution equation \cite{Hautmann2017,Hautmann2018}, which can be combined with finite-order matrix elements either by matching \cite{Martinez2019a,Martinez2020}, or merging \cite{Martinez2021,Martinez2022}. The transverse-momentum distribution of the $J/\psi$ particle has already been studied with the event generator CASCADE using more traditional $k_t$-factorization calculations, e.g., with LO off-shell matrix elements and the color-singlet model, see, for instance, Ref. \cite{Jung2011}.\\

 In Eq.~(\ref{ktfac}), the differential cross section $\hat{\sigma}_{ab\to Q\Bar{Q}}$ for the production of an unpolarized heavy-quark pair is computed at NLO with MG5, option -p, which includes all necessary subtraction terms. Option -p means the parton showers are not performed within MG5, but with other tools, in our case, CASCADE3 \cite{Jung2010,Baranov2021} and pythia 6. We worked in a 3-flavor scheme with $m_c=1.3$ GeV, and used the CT14nlo\_NF3 \cite{Dulat2016} PDFs. This choice is imposed by the fact that, in MG5, the variable-flavor-number scheme (VFNS) works with $m_c=0$ (i.e., it is a Zero-Mass VFNS). In \cite{Thorne2008}, the authors argue that, in practice, it is generally better\footnote{A case where this statement does not apply is heavy-quark production at leading order (LO). Here, VFNS PDFs with a 3-flavor scheme cross section strongly underestimate the data. This is because the partonic cross section for the $cg\to cg$ process is numerically large. While it is already included at LO in a VFNS, it appears only at NLO in a 3-flavor scheme.} to use VFNS PDFs, even with cross sections computed in a 3-flavor scheme. We checked that changing the CT14nlo\_NF3 for the CT14nlo gives similar results. With the different PDFs sets tested in our study, we observed a negligible  or small variation of the final result compared to scale variation. The function $d_{Q\bar{Q}}(\mu,\mu_0)$ implements the evolution of the heavy-quark pair, with a convolution on the variable $z$ representing the momentum fraction of the initial $Q\bar{Q}$ state carried by the final quarkonium. To be clear, we do not have an analytical expression for the evolution function $d_{Q\bar{Q}}$. All our calculations are performed by event generators using distribution probabilities, with the three steps being: 1. fixed-order production with MG5, 2. evolution with a timelike cascade, and 3. hadronization with the CEM. The latter is applied at $\mu_0\sim m_{J/\psi}$ and sums all pairs with an invariant mass between $2m_c$ and $2m_{D^0}$~\cite{Fritzsch1977,Halzen1977}
\begin{equation}
    \frac{d\sigma}{dp_t}=F_{J/\psi}\int_{2m_c}^{2m_{D^0}}\frac{d\sigma^{Q\Bar{Q}}}{dm_{Q\Bar{Q}}dp_t}dm_{Q\Bar{Q}}. \label{eqcem}
\end{equation}
We fixed the parameter $F_{J/\psi}$ to the value found in Ref.~\cite{Lansberg2020} at NLO\footnote{$\mathcal{O}(\alpha_s^3)$ contributions are considered to be LO in \cite{Lansberg2020} because the $\mathcal{O}(\alpha_s^2)$ diagrams do not contribute to $p_t>0$ in collinear factorization. This is not the case when using the $k_t$-factorization, and $\mathcal{O}(\alpha_s^3)$ contributions are NLO.}:
\begin{equation}
    F_{J/\psi}=0.014.
\end{equation}
In that sense, our calculations are parameter free. As visible from Eq.~(\ref{ktfac}), our calculations include only the direct contribution, and will be compared to prompt-$J/\psi$ data. Finally, note that we did not use the improved CEM \cite{Ma2016}
\begin{equation}
  \frac{d\sigma}{d^3p}=F_{J/\psi}\int_{m_{J/\psi}}^{2m_{D^0}}\frac{d\sigma^{Q\Bar{Q}}}{dm_{Q\Bar{Q}}dp^3_{Q\Bar{Q}}}\delta^3\left(p-\frac{m_{J/\psi}}{m_{Q\Bar{Q}}}p_{Q\Bar{Q}} \right)dm_{Q\Bar{Q}},
\end{equation}
because the reduced phase space ($m_{J/\psi}>2m_c$) requires more statistics. Note that the difference between the CEM and ICEM is visible for $p_t \lesssim 15$ GeV \cite{Lansberg2020a}. Several studies using LO off-shell matrix elements and the CEM are also available~\cite{Cheung2018,Maciula2019,Chernyshev2022}.\\

Contrary to the factorization formalism, the CEM does not organize the calculation into LP and NLP contributions. NLO diagrams with a gluon splitting into a $Q\bar{Q}$ pair (see Fig.~\ref{nloctr}) contribute to both LP and NLP \cite{Kang2012, Kang2015}. In the factorization formalism, the LP contribution is removed by a subtraction term, avoiding mass singularities and double counting with the LP contribution $\sigma_f\otimes D_{f\to H}$. The latter describes the short-distance production of a parton $f$ followed by its fragmentation into the quarkonium $H$. In our calculations with MG5, no double counting occur because the contribution $\sigma_f\otimes D_{f\to H}$ is not included (then, several LP contributions are missing). Moreover, we keep the mass of heavy quark in the cross section, so the gluon cannot get on-shell and does not produce mass singularities. Then, CEM calculations do not include the subtraction term\footnote{The subtraction term mentioned in this paragraph is different from the subtraction terms in MG5.} and retain some LP contributions, which are responsible for the overestimation of the cross section by fixed-order calculations. The missing LP contributions have certainly an effect on the determination of $F_{J/\psi}$, but not on the shape of the distribution at intermediate and large transverse momentum.\\

In the subsequent section, we compare our calculations with high-energy data, with the expectation that the evolution should improve the results obtained by fixed-order calculations. Indeed, a consequence of the evolution is energy loss by the heavy-quark pair, shifting fixed-order calculations of Fig.~\ref{madfo} to the left. The cascade usually starts at $\mu_c \simeq \mu \simeq p_t$, and we chose $\mu_c=\mu$ by simplicity. To larger $p_t$ corresponds a longer evolution and a larger shift in $p_t$, making the theoretical prediction softer. From a more theoretical point of view, we will obtain a softer spectrum because the evolution modifies the FO behavior of $1/p_t^4$ to \cite{Kang2014}
\begin{equation}
    \frac{m_Q^2}{p_t^4\mu^2}.
\end{equation}

\section{$J/\psi$ production at small, intermediate and large transverse momentum \label{secRes}}
\begin{figure}[!h]
\begin{center}
 \includegraphics[width=22pc]{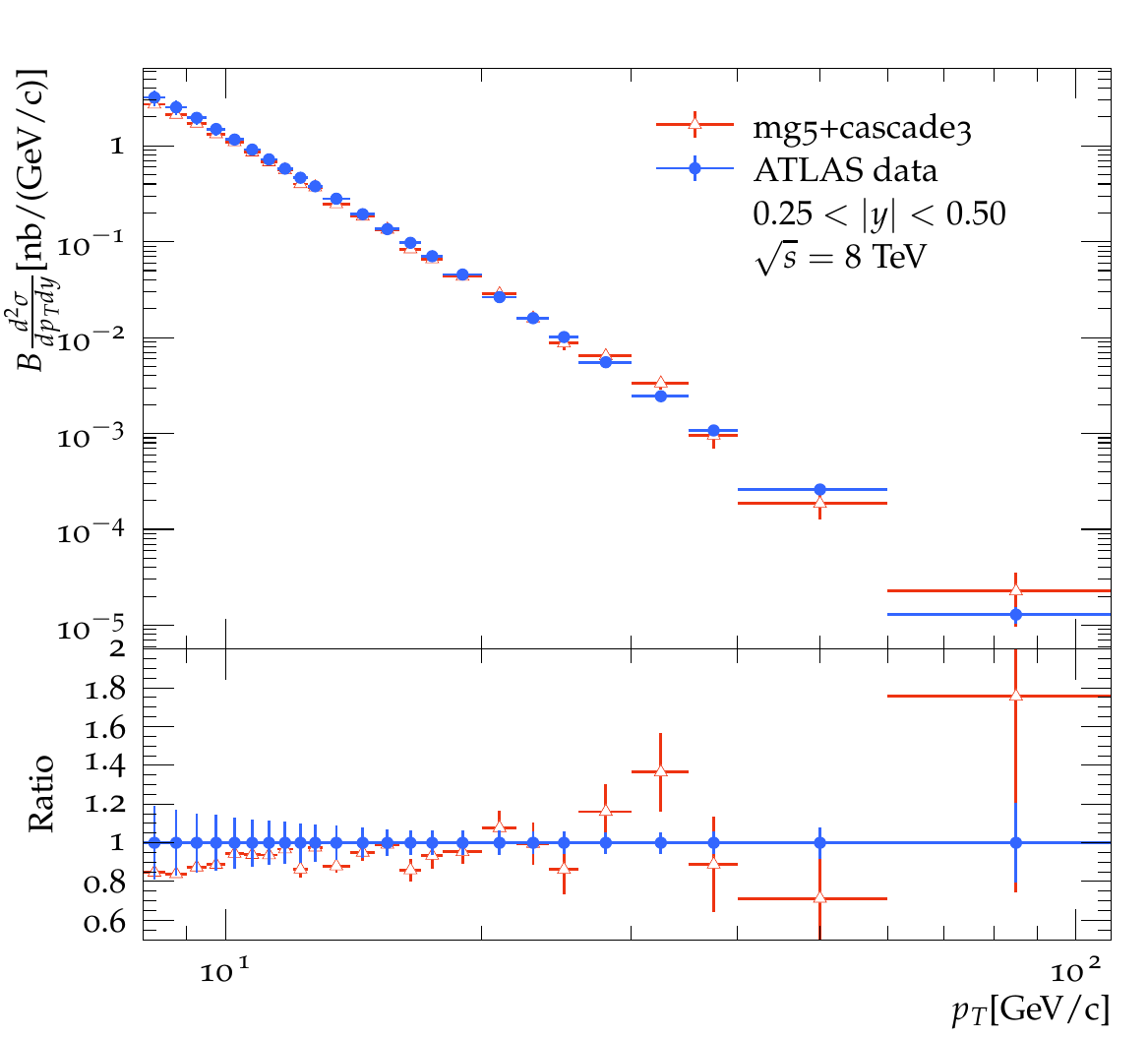}
\end{center}
\caption{\label{fullhpt} Comparison of the prompt-$J/\psi$ differential cross section measured by ATLAS \cite{Aad2016} as a function of transverse momentum $p_T$ (blue dots) with our calculations Eqs.~(\ref{ktfac}) and (\ref{eqcem}) (red triangles). The mg5+cascade3 error bars includes only the statistical uncertainty. The ratio theory/exp is displayed in the bottom panel.}
\end{figure}
We start with ATLAS data on prompt $J/\psi$ at 8 TeV \cite{Aad2016}, already compared to the CEM at NLO (fixed order) in Fig.~\ref{nloexp}, see also Fig. 33 of Ref. \cite{Lansberg2020}. While fixed-order calculations fail, we observe in Fig.~\ref{fullhpt} a good agreement between our full calculations (mg5+cascade3) and data. The results obtained for the other rapidity ranges are displayed at the end of this manuscript. At large $p_t$, the main difference between Figs.~\ref{madfo} and \ref{fullhpt} is the timelike cascade. As expected, including the $Q\bar{Q}$ evolution improved the result. It was less clear, however, that this sole effect would bring CEM calculations in agreement with data, this model being quite simple. But we should be careful to not overinterpret these results. Our calculations do not include the feed down contributions, and the timelike cascade is only an approximation of the true evolution equation. Still, improving these points will certainly not change our conclusion. Note that, additionally of the energy lost by the heavy-quark pair, another effect of the timelike cascade is to change the invariant mass, sometimes destroying quarkonium candidates with $2m_c < m_{Q\bar{Q}} < 2m_{D^0}$. The opposite situation is also possible, promoting $Q\bar{Q}$ pairs produced at fixed order with the incorrect invariant mass to quarkonium candidates. An advantage of event generators is to automatically include such kinematical effects.
\begin{figure}[!h]
\begin{center}
 \includegraphics[width=22pc]{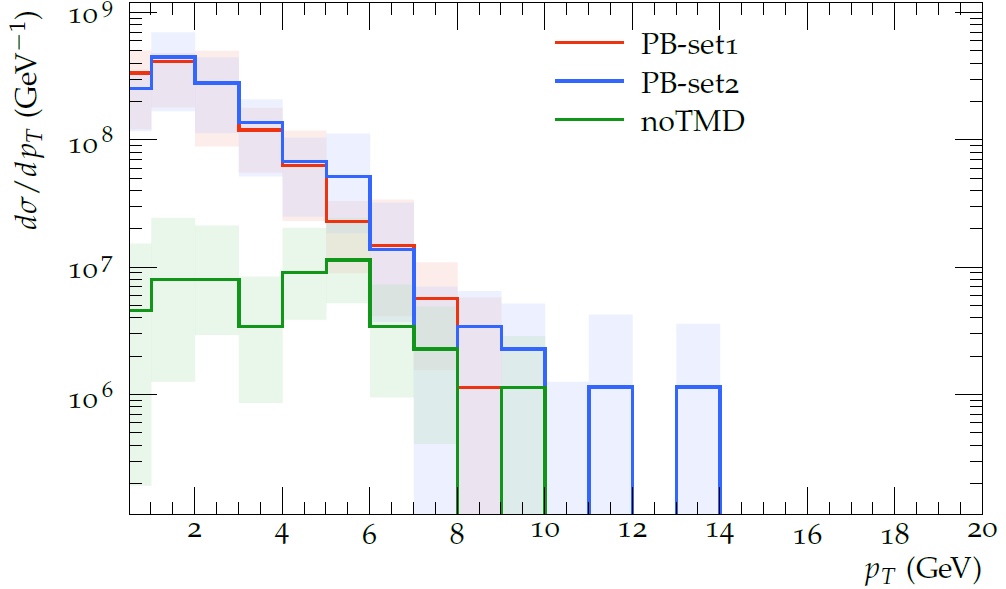}
\end{center}
\caption{\label{compTMD} Differential cross section, Eq.~(\ref{ktfac}) (without the last factor), at $\sqrt{s}=200$ GeV as a function of the $Q\bar{Q}$ transverse momentum. The red and blue lines correspond to the UPDFs PB set 1 and 2. The green line corresponds to the case with no initial $k_t$.}
\end{figure}

While taking into account the initial transverse momentum is not primordial at large $p_t$ (with our formalism), this effect is essential at small $p_t$. This is illustrated in Fig.~\ref{compTMD}, where we plot the result obtained at $\sqrt{s}=200$ GeV, with the initial transverse momentum obtained from the PB set1 and set2, and with no initial transverse momentum in green. While both PB sets give a similar result, the green line corresponding to no initial transverse momentum is more than one order of magnitud below. Since the LO contribution is exactly zero in the case of no initial $k_t$, this line corresponds to the pure NLO contribution. We observe that this contribution starts to dominate at $p_t\sim 10$ GeV.

\begin{figure}[!h]
\begin{center}
 \includegraphics[width=23pc]{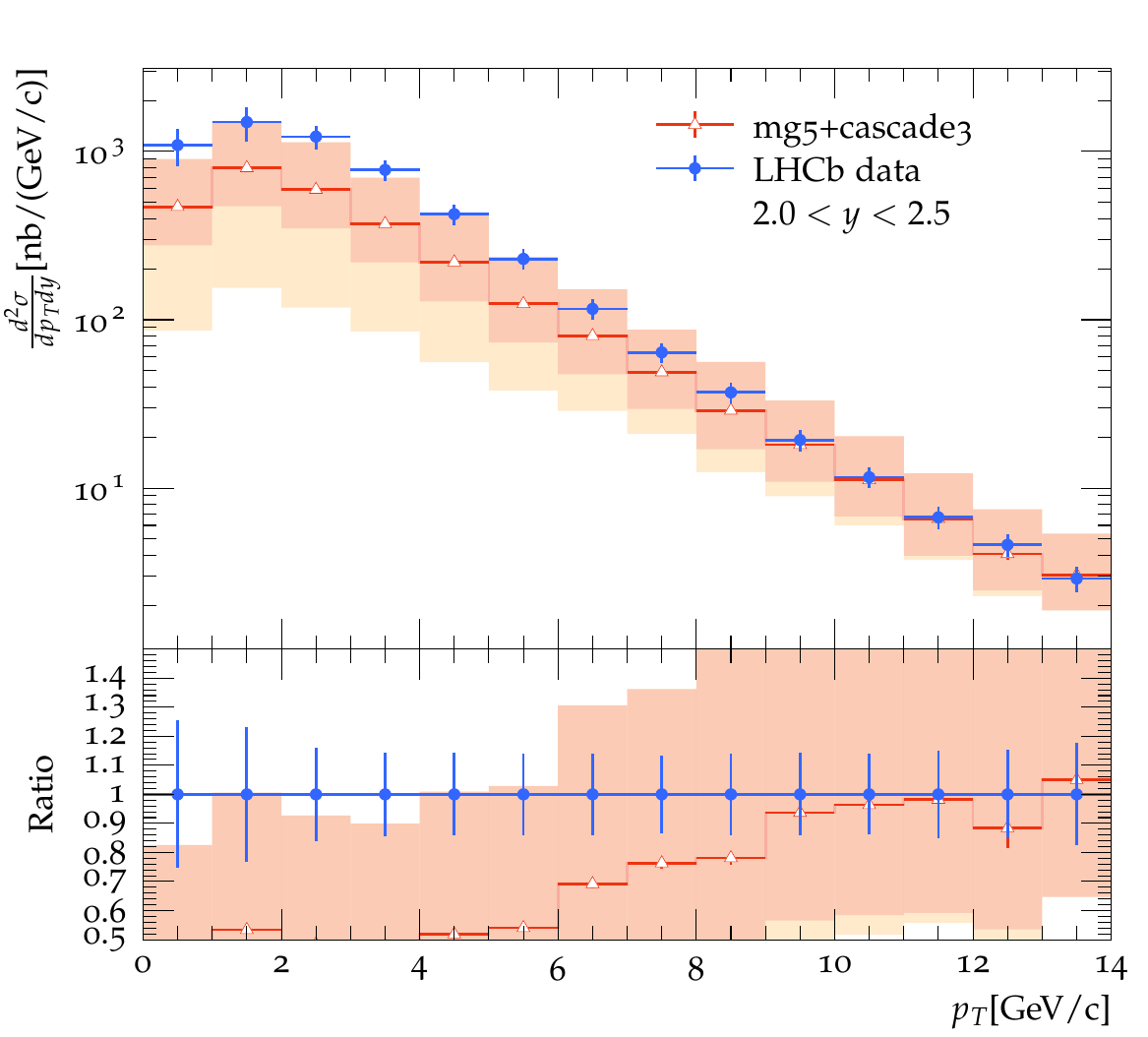}
\end{center}
\caption{\label{lhcb7_20}  Comparison of the $J/\psi$ differential cross section measured by the LHCb collaboration at 7 TeV~\cite{LHCbQ2011} (blue dots) with our calculations Eqs.~(\ref{ktfac}) and (\ref{eqcem}) (red triangles). The mg5+cascade3 error bands include statistical and scale uncertainties, see the text  below Fig.~\ref{compTMD} for more details. The ratio theory/exp is displayed in the bottom panel.}
\end{figure}

\begin{figure}[!h]
\begin{center}
 \includegraphics[width=24pc]{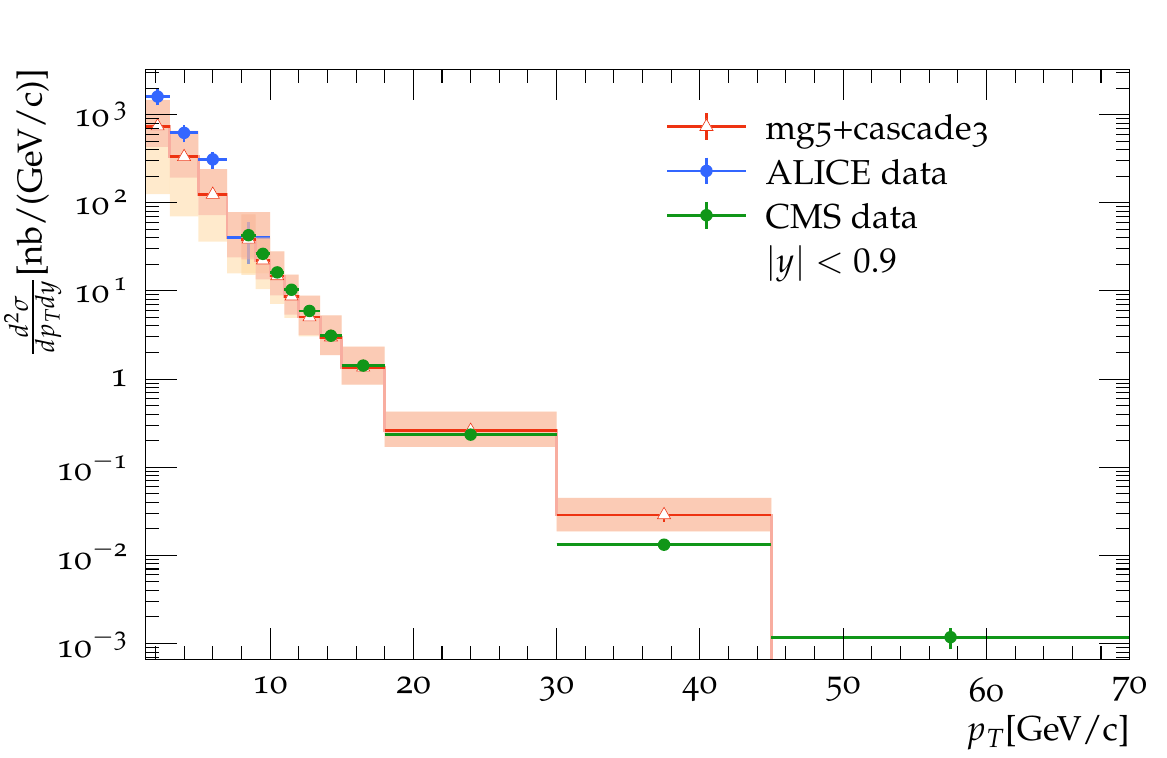}
\end{center}
\caption{\label{AlCms} Comparison of the 7 TeV $J/\psi$ differential cross section measured by ALICE \cite{AliceQ2012} (blue dots) and CMS \cite{Chatrchyan2012} (green dots)  with our calculations Eqs.~(\ref{ktfac}) and (\ref{eqcem}) (red triangles). There is no theoretical prediction in the last bin because of statistics. The mg5+cascade3 error bands include statistical and scale uncertainties, see the text below Fig.~\ref{compTMD} for more details.}
\end{figure}

In Figs.~\ref{lhcb7_20} and \ref{AlCms}, we compare our calculations at 7 TeV with LHCb \cite{LHCbQ2011}, ALICE \cite{AliceQ2012}, and CMS \cite{Chatrchyan2012} data.
As discussed in more detail in section \ref{secUni}, we worked with the same value of the non-perturbative parameter $F_{J/\psi}$. Our central values underestimate data at $p_t \lesssim 5$ GeV but are still in agreement within uncertainties. The theoretical uncertainties, shown as red and beige color bands, include the variation of factorization and/or renormalization scales, as well as statistical uncertainties. We obtained the red color band by varying the  factorization scale between $m_t/2$ and $2m_t$, with $m_t$ the $J/\psi$ transverse mass, while the beige color band is built from the conventional 7-point variation\footnote{In the 7-point variation, we vary independently the factorization and renormalization scales between $m_t/2$ and $2m_t$ leading to nine combinations, from lower ($\mu=m_t/2$), central ($\mu=m_t$) and upper ($\mu=2m_t$) limits. Then, the two combinations leading to the highest and lowest variation with respect to central values are ignored.}. Both techniques give the same upper limit, but the 7-point variation gives a wider error band.\\

Let us come back for a moment to our description of low $p_t$ data. In Sec.~\ref{secForm}, we mentioned that our calculations miss some LPs. It is true for any calculation which does not include the fragmentation contribution $\sigma_f\otimes D_{f\to H}$, in particular, for the NRQCD cross section (\ref{nrqcd}). By definition, LP contributions produce the same distribution and are dominant at large $p_t$. Then, the missing LPs can be compensated by an appropriate (larger) choice of the parameter $F_{J/\psi}$, which explains our good description of ATLAS data. But then, one could expect an overestimation of low-$p_t$ data where the NLPs are dominant. It is clearly not the case. One hypothesis is that the missing LPs do not contribute significantly to the cross section, at least in the transverse momentum range of Fig.~\ref{fullhpt}. Another explanation is that the CEM at NLO misses important contributions at low $p_t$, for instance, higher-order contributions or the fragmentation of a gluon emitted during the spacelike cascade into a quarkonium. It would be interesting to see a dedicated study on the role of these different contributions.\\

Meanwhile, we conclude that the phenomenological CEM, with the evolution of the heavy-quark pair included, can describe data on the whole $p_t$ range shown in this study. Again, our main goal is not to defend the CEM, but to show the effect of scale evolution in a simple case. It could be interesting test the CEM further, with less inclusive observables such as the $z$-distribution of a $J/\psi$ within a jet. Fragmentation contributions matched with NRQCD have been compared to this observable, see, for instance, Refs. \cite{Bain2017,Baumgart2014}.

\section{Universality of $F_{J/\psi}$ \label{secUni}}
\begin{figure}[!h]
\begin{center}
 \includegraphics[width=24pc]{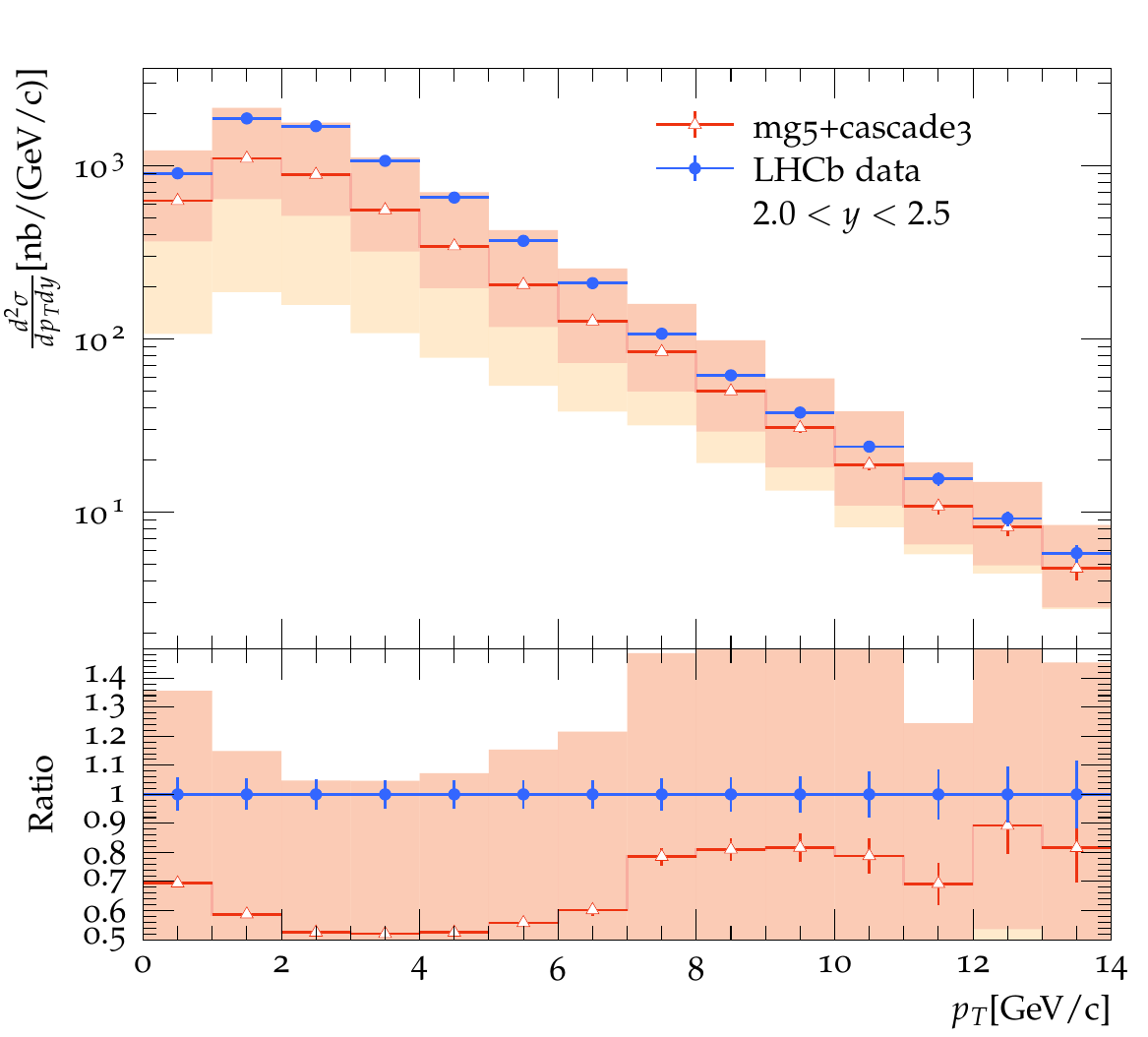}
\end{center}
\caption{Comparison of the $J/\psi$ differential cross section measured by the LHCb collaboration at 13 TeV \cite{Aaij2015, Aaij2017} (blue dots) with our calculations Eqs.~(\ref{ktfac}) and (\ref{eqcem}) (red triangles). The mg5+cascade3 error bands include statistical and scale uncertainties, see the text  below Fig.~\ref{compTMD} for more details. The ratio theory/exp is displayed in the bottom panel.\label{figlhcb13}}
\end{figure}

\begin{figure}[!h]
\begin{center}
 \includegraphics[width=24pc]{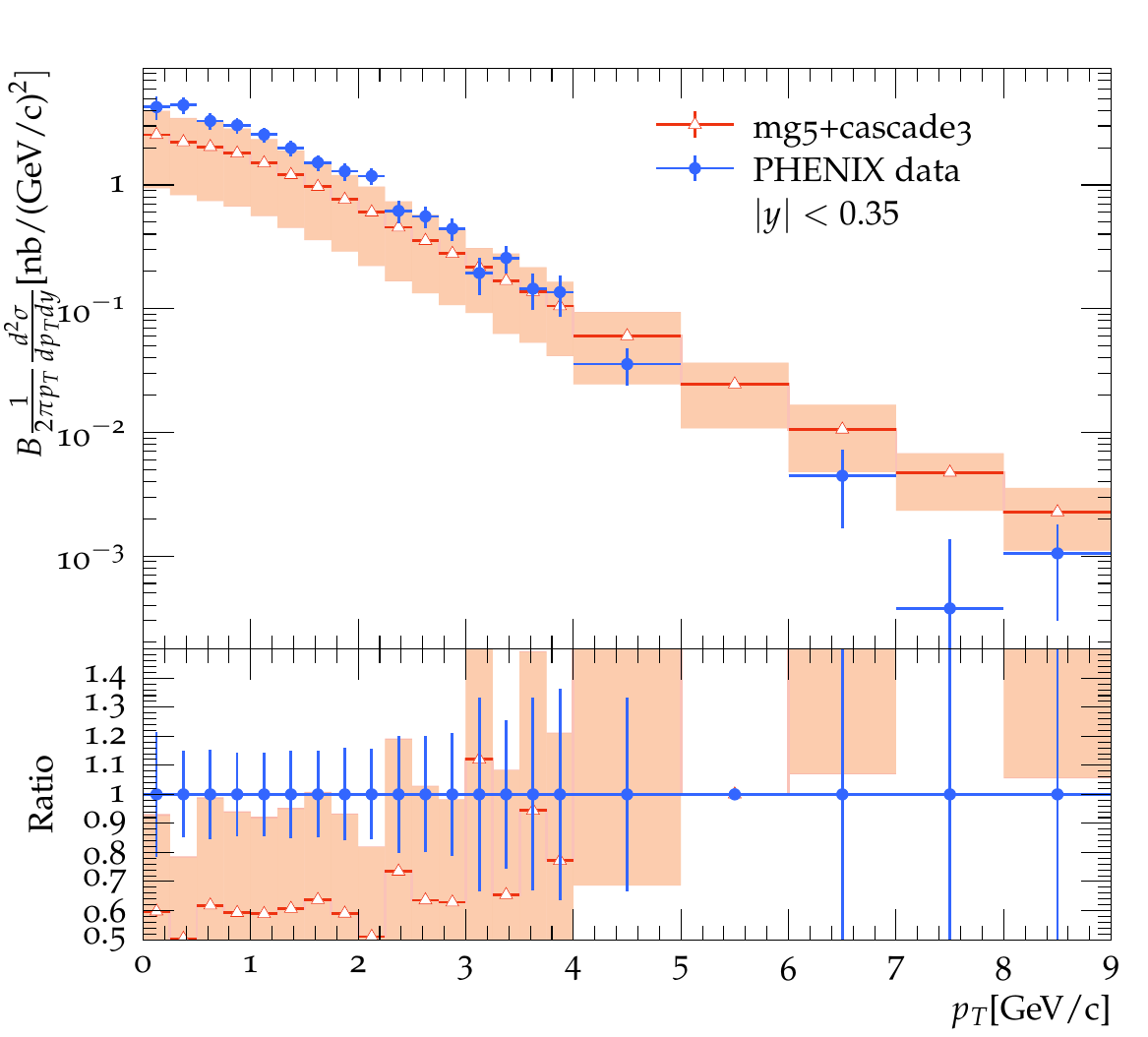}
\end{center}
\caption{\label{plotPhenix}Comparison of the $J/\psi$ differential cross section measured by the PHENIX collaboration at 200 GeV \cite{Adare2007} (blue dots)  with our calculations Eqs.~(\ref{ktfac}) and (\ref{eqcem}) (red triangles). The mg5+cascade3 error band includes statistical and scale uncertainties, see the text  below Fig.~\ref{compTMD} for more details. The ratio theory/exp is displayed in the bottom panel.}
\end{figure}

In Figs.~\ref{figlhcb13} and \ref{plotPhenix}, we compare our calculations to LHCb data at 13 TeV \cite{Aaij2015, Aaij2017} and PHENIX data at 200 GeV \cite{Adare2007}. We used the same value of the non-perturbative parameter, i.e., $F_{J/\psi}=0.014$, and obtained results similar to those of Sec.~\ref{secRes}.

We conclude on the universality of $F_{J/\psi}$, at least on this range of energies. On the opposite, using the $k_t$-factorization with LO off-shell matrix elements, Ref.~\cite{Chernyshev2022} reaches a different conclusion, with $F_{J/\psi}(200 \text{ GeV})$ about 10 times larger than at 7 TeV (except for the ALICE experiment, with a factor of 2.7). The formalisms used in \cite{Chernyshev2022} being quite different, it is hard to pinpoint the precise reason explaining this difference. However, concentrating on small and intermediate $p_t$, we can make the following comments. This kinematical region, being dominated by the LO contribution, it seems unlikely that the difference between Ref.~\cite{Chernyshev2022} and the present work can be explained by the fact that we included the NLO contribution. Another difference is the use of an on-shell cross section instead of an off-shell one. However, these two quantities are supposed to be close at small $k_t$.\\

Another possible explanation is the $k_t$ dependence of the KMRW UPDFs \cite{Kimber2001,Watt2003}, already discussed in \cite{Guiot2020a,Guiot2021,Guiot2023} in the context of D-meson production, see also \cite{Guiot2019,Hautmann2019}. These UPDFs, used in \cite{Chernyshev2022}, are constrained for $k_t\in [0,\mu]$ by their relation to collinear PDFs
\begin{equation}
    f_{k/h}(x;\mu)=\int_0^{\mu^2}F_{k/h}(x,k_t;\mu)dk_t^2.
\end{equation}
Here $f_{k/h}$ gives the collinear distribution of a parton of flavor $k$ in the hadron $h$. On the opposite, the cross section, Eq.~(\ref{ktfac}), is integrated up to $k_t^2 \sim s$. At fixed $\mu$, or equivalently fixed $p_t$, and increasing $s$, the unconstrained part of the UPDFs, $k_t\in [\mu, \sqrt{s}]$, contributing to the cross section increases. To compensate that, we expect a decrease of $F_{J/\psi}$, which acts as a normalization factor. This decrease of $F_{J/\psi}$ with energy is indeed observed in Ref.~\cite{Chernyshev2022}. On the opposite, the PB UPDFs used in our study are constrained on the full phase space
\begin{equation}
    f_{k/h}(x;\mu)=\int_0^{\infty}\frac{F_{k/h}(x,{\bf k}_t;\mu)}{\pi}d^2{\bf k}_t,
\end{equation}
and the numerical value of these functions in the region $[\mu, \sqrt{s}]$ is negligible. Then, the $k_t$ integral of Eq.~(\ref{ktfac}) is effectively cut off at $k_t \sim \mu$, resulting in a energy independent $F_{J/\psi}$.\\

Whatever the correct explanation is, it shows the potential of quarkonia data to constrain the (gluon) UPDFs.

\section{Comparison with EPOS4 \label{secepos}}
\subsection{Short presentation of the formalism}
The EPOS4 project is an attempt to construct a realistic model for
describing relativistic collisions of different systems, from proton-proton
(pp) to nucleus-nucleus (AA), at energies from several TeV per nucleon
down to several GeV, see Ref.~\cite{Werner2023a} and references therein. In this section, we discuss the relevant features of this event generator for $J/\psi$ production.\\

\begin{figure}[!h]
\noindent \begin{centering}
$\qquad$(a)\hspace*{4cm}(b)\hspace*{4cm}$\qquad$\\
\includegraphics[scale=0.25]{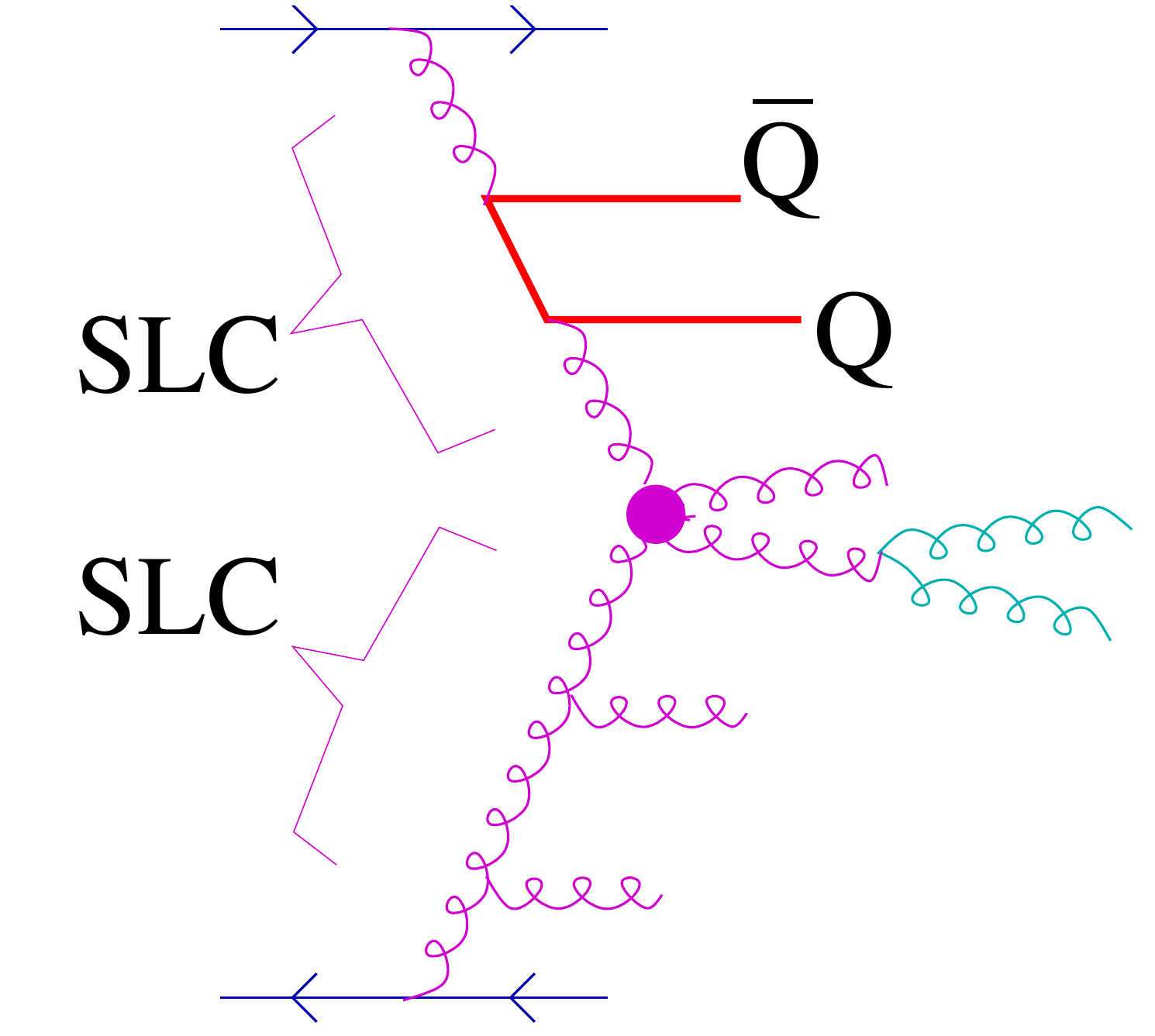}$\qquad$\includegraphics[scale=0.25]{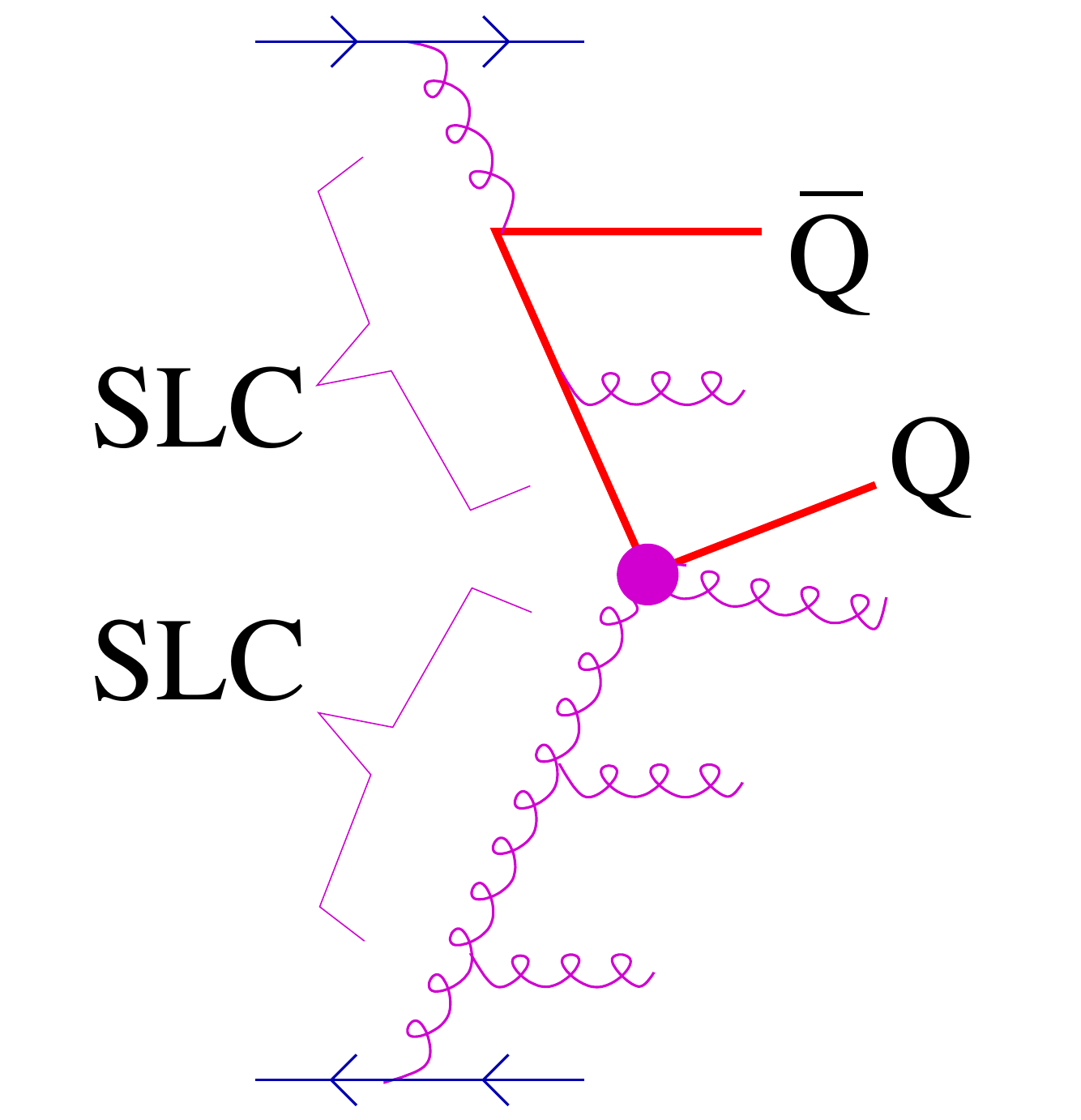}\\
$\qquad$\hspace*{2cm}(c)\hspace*{7cm}$\qquad$\\
$\qquad$\includegraphics[scale=0.25]{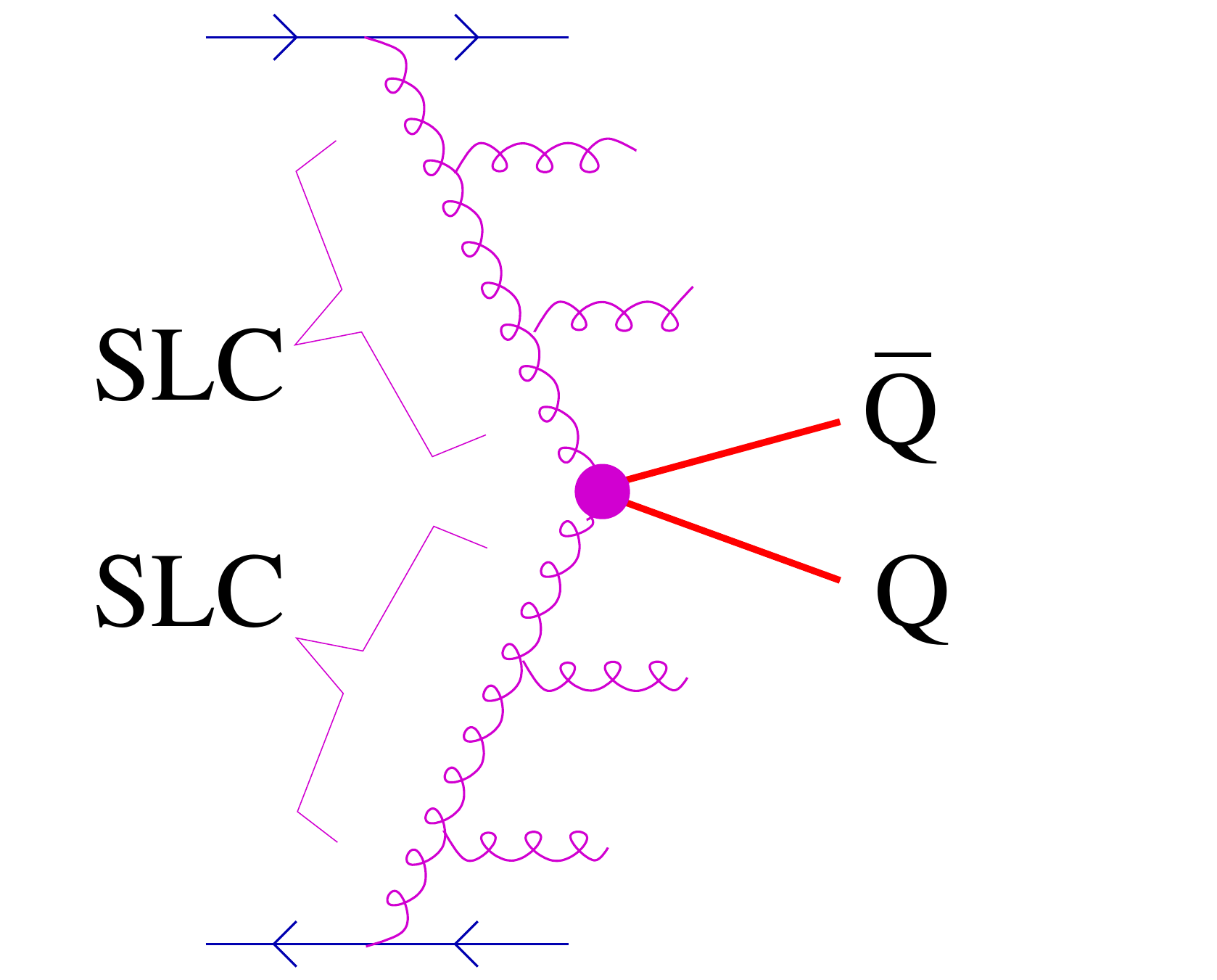}\\
\par\end{centering}
\noindent \centering{}$\qquad$(d)\hspace*{4cm}(e)\hspace*{4cm}$\qquad$\\
\includegraphics[scale=0.25]{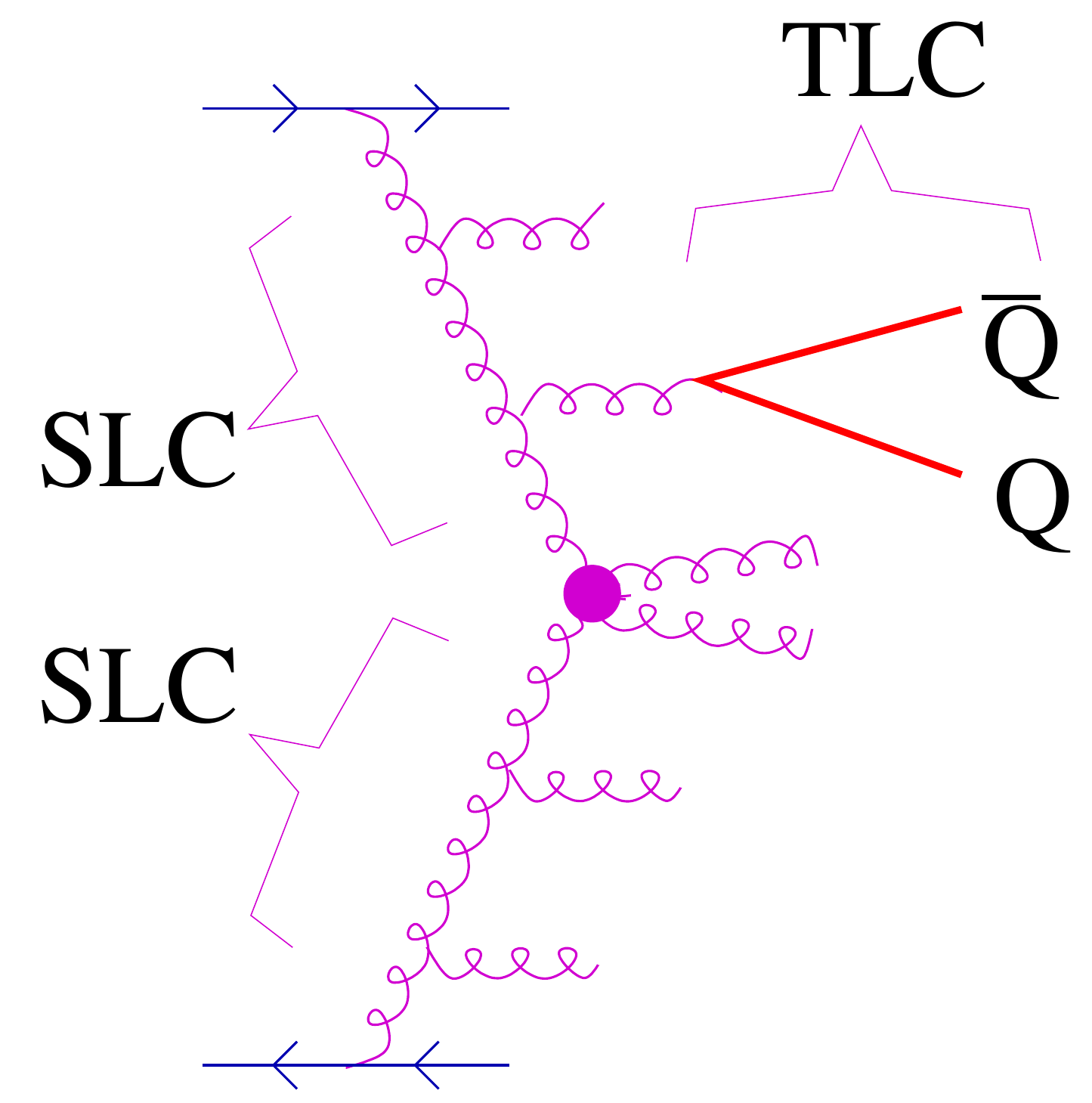}$\qquad$\includegraphics[scale=0.25]{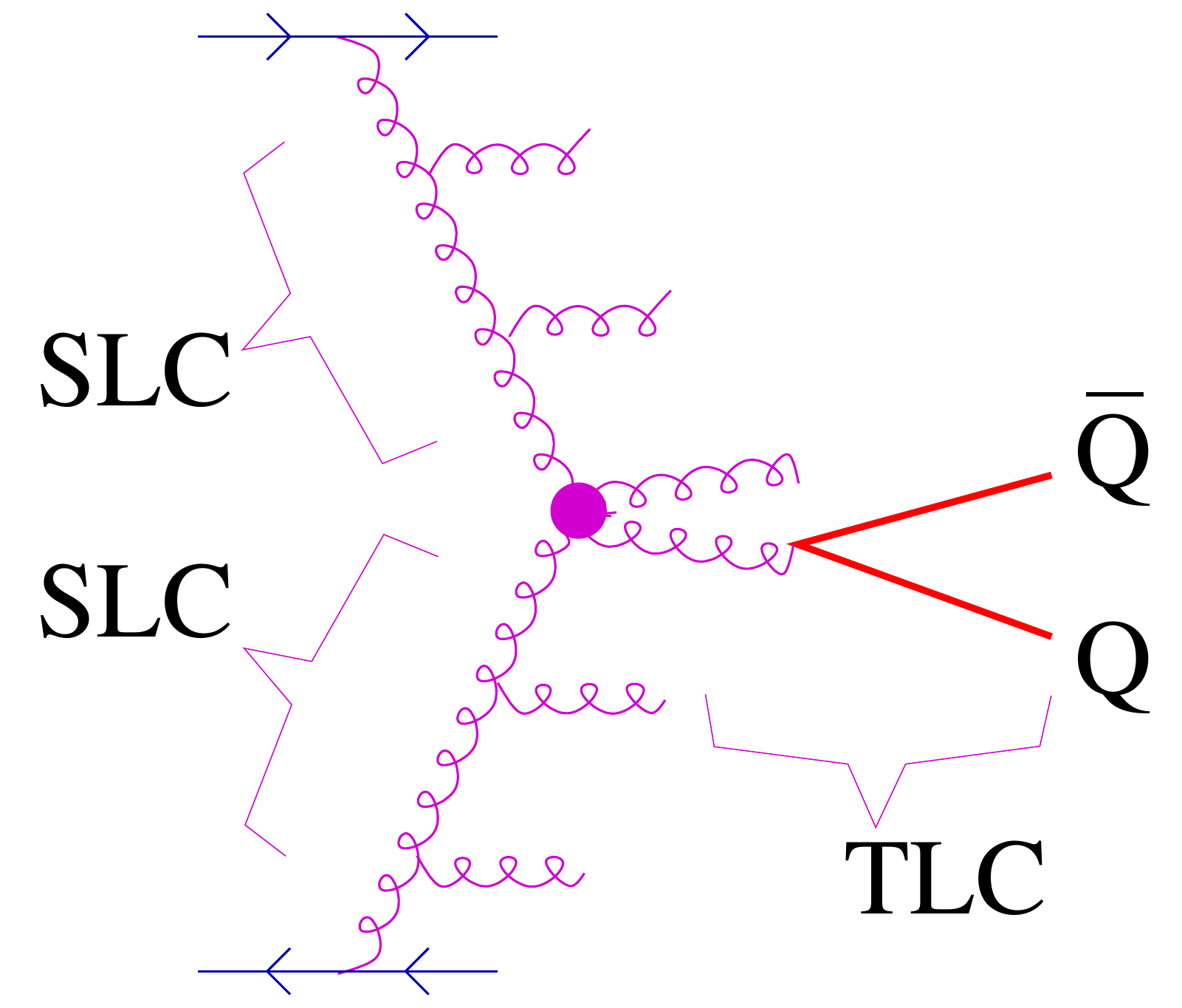}\caption{Heavy flavor production in parton ladders (cut pomeron). SLC refers to spacelike cascade, and TLC to timelike cascade. Even if we show mainly gluonic lines, we use a GM-VFN scheme and similar diagrams with quark lines can also be drawn.\label{charm-1}}
\end{figure}

EPOS4 is based on parallel scatterings, with the initial nucleons
(and their partonic constituents) being involved, happening instantaneously at very high energies. The theoretical tool is S-matrix theory,
using a particular form of the proton-proton scattering S-matrix (Gribov-Regge (GR) approach \cite{Gribov:1967vfb,Gribov:1968jf,GribovLipatov:1972,Abramovsky:1973fm}),
which can be straightforwardly generalized to be used for nucleus-nucleus
collisions. This S-matrix approach has a modular structure
based on so-called ``cut Pomerons'', representing inelastic parton-parton
scatterings. Each cut pomeron has the ladder structure shown in Fig.~\ref{charm-1}. 
Two initial spacelike partons evolve according to the DGLAP equation \cite{GribovLipatov:1972,AltarelliParisi:1977,Dokshitzer:1977}, with the center of the ladder corresponding to a leading-order $2\to 2$ cross section. The timelike partons emitted during the spacelike evolution and those produced in the $2\to 2$ scattering will also evolve according to the DGLAP equation: this is the timelike cascade discussed earlier. In Fig.~\ref{charm-1}, the heavy quark and antiquark are represented by red lines. We should stop here to compare the different formalisms.  Diagrams (a), (b), and (d) are not included by the MG5+cascade3 study. On the other hand, the $gg\to gQ\bar{Q}$ diagram in the center of cut pomeron (e) is taken into account by the NLO cross section provided by MG5. However, EPOS4 works with LO cross section, and the splitting of the gluon into the heavy-quark pair happens in the timelike cascade. It is worth noting that the $Q\bar{Q}$ pair can be produced at any step during the timelike cascade, and that all kind of timelike partons can be produced in the $2\to 2$ scattering. Consequently, and contrary to our MG5+cascade3 calculations, EPOS4 includes contributions similar to the LP contributions $\sigma_f\otimes D_{f\to H}$ of the factorization formalism. Finally, the factorization formalism and our MG5+cascade3 calculations do not include, for instance, diagram (a), responsible for the production of low-$p_t$ $J/\psi$.\\

Once the timelike cascade ends, we compute the $J/\psi$ yield according to the CEM: we sum over all $c-\bar{c}$ pairs, compute their masses $m=\sqrt{p(c)\cdot p(\bar{c})}$, and count them with probability $F_{J/\psi}=0.028$ , in case of $m_{J/\psi}<m<2m_{D}$. There is no particular reason for using $m_{J/\psi}$ rather than $2m_c$, and it affects only the value of $F_{J/\psi}$. In the next version of EPOS4, we will use $2m_C$.\\

When developing matrix elements in terms of multiple scattering diagrams, the large majority of the diagrams cancel when it comes to inclusive cross sections. This is the Abramovsky-Gribov-Kancheli cancellations \cite{Abramovsky:1973fm}. The challenge for EPOS4 is to use the full parallel multiple scattering scenario, but in such a way that for inclusive cross section the cancellations actually work. This is the new part in EPOS4, strongly based on an interplay between parallel scatterings and saturation. It is discussed in a separate publication \cite{Werner2023}. But EPOS4 keeps all contributions, and goes beyond inclusive cross sections. This is important, for instance, when studying high multiplicity events, or for $J/\psi$ pair production. We plan to study the latter in an separate work.

\subsection{Results}
\begin{figure}[!h]
\begin{center}
 \includegraphics[width=22pc]{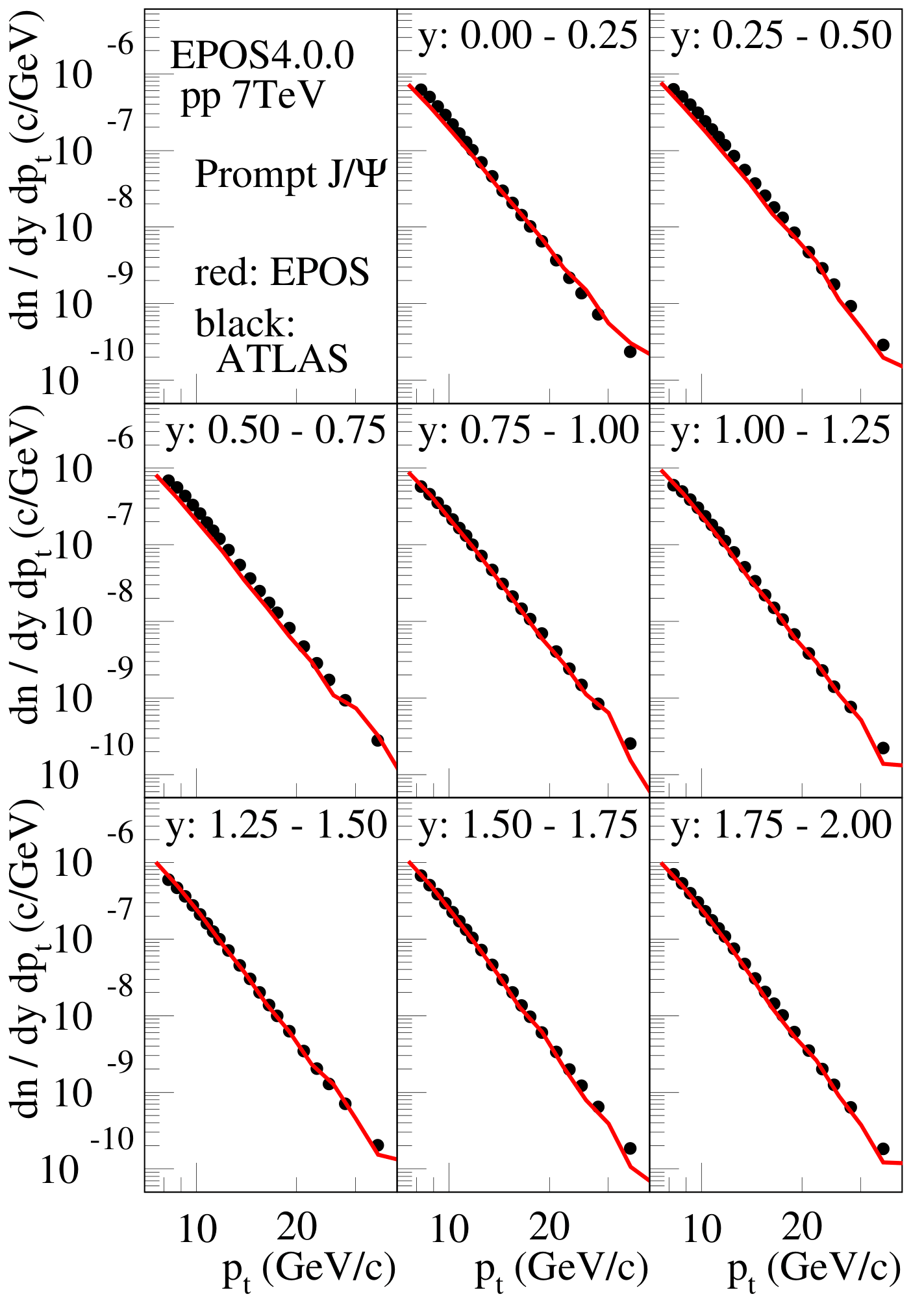}
\end{center}
\caption{Comparison of the prompt-$J/\psi$ differential cross section measured by ATLAS \cite{Aad2016} as a function of transverse momentum (black dots) with EPOS4 simulations in red. The rapidity range is indicated in the top of each plot.\label{eposatlas}}
\end{figure}

The two main similarities between EPOS4 and our mg5+cascade3 calculations are the use of a timelike cascade, and the hadronization of the produced $Q\bar{Q}$ pairs based on the CEM. Fig.~\ref{eposatlas} shows that EPOS4 results for ATLAS data are qualitatively the same as mg5+cascade3 calculations.
There are, however, several relevant differences between the two formalisms:
\begin{itemize}
    \item EPOS4 uses LO cross sections, and the madgraph NLO contribution is taken into account by the timelike cascade.
    \item It includes all LP contributions. Indeed, EPOS4 produces any kind of partons which evolve with the timelike cascade, and can split into a $Q\bar{Q}$ pair at any time.
    \item EPOS4 works with a GM-variable-flavor scheme, with $m_c = 1.27$ GeV. It implies that contributions similar to graph (b) of Fig.~\ref{charm-1} are included, while absent for mg5+cascade3.
    \item The details of the implementation of the timelike cascade may be different.
\end{itemize}
Despite these differences, both formalisms show similar results, in relatively good agreement with data, and far from the result obtained with NLO fixed-order calculations, Fig.~\ref{madfo}. The key ingredient is the $Q\bar{Q}$ evolution.\\ 

Still, one may be surprised by the fact that both calculations give similar results for ATLAS data, where the LP contributions dominate. While EPOS4 allows a produced gluon to emit, in principle, any number $n$ of partons before splitting into the heavy-quark pair, our calculations with MG5 include only the case $n=0$. The explanation is probably that the produced $Q\bar{Q}$ pair keeps emitting partons, mainly gluons. Then, the global picture is that a color object propagates, emitting partons, and we find a heavy-quark pair at the end of the process. Qualitatively, it does not really matter if the partons are emitted by a gluon or the heavy-quark pair\footnote{Of course, the rate of emissions is not exactly the same.}. The final result is energy loss making the spectrum softer. However, as already discussed in Sec.~\ref{secRes}, the missing LP contributions in our mg5+cascade3 calculations could affect the value of $F_{J/\psi}$.

Finally, we remark that the probability for the propagating gluon to emit a large number $n$ of gluons before producing the $Q\bar{Q}$ pair is suppressed because the virtuality $Q^2$ of timelike partons decreases after each emission. Creating a $Q\bar{Q}$ pair requires at least a virtuality of $Q^2_{\text{min}}=4m_Q^2$, limiting the number of emissions before its production.

\section{Complementary figures: Comparison with ATLAS data}
In this section, all figures are the same as Fig.~\ref{fullhpt}, but for different rapidity ranges.
\begin{figure}[!h]
\begin{center}
 \includegraphics[width=22pc]{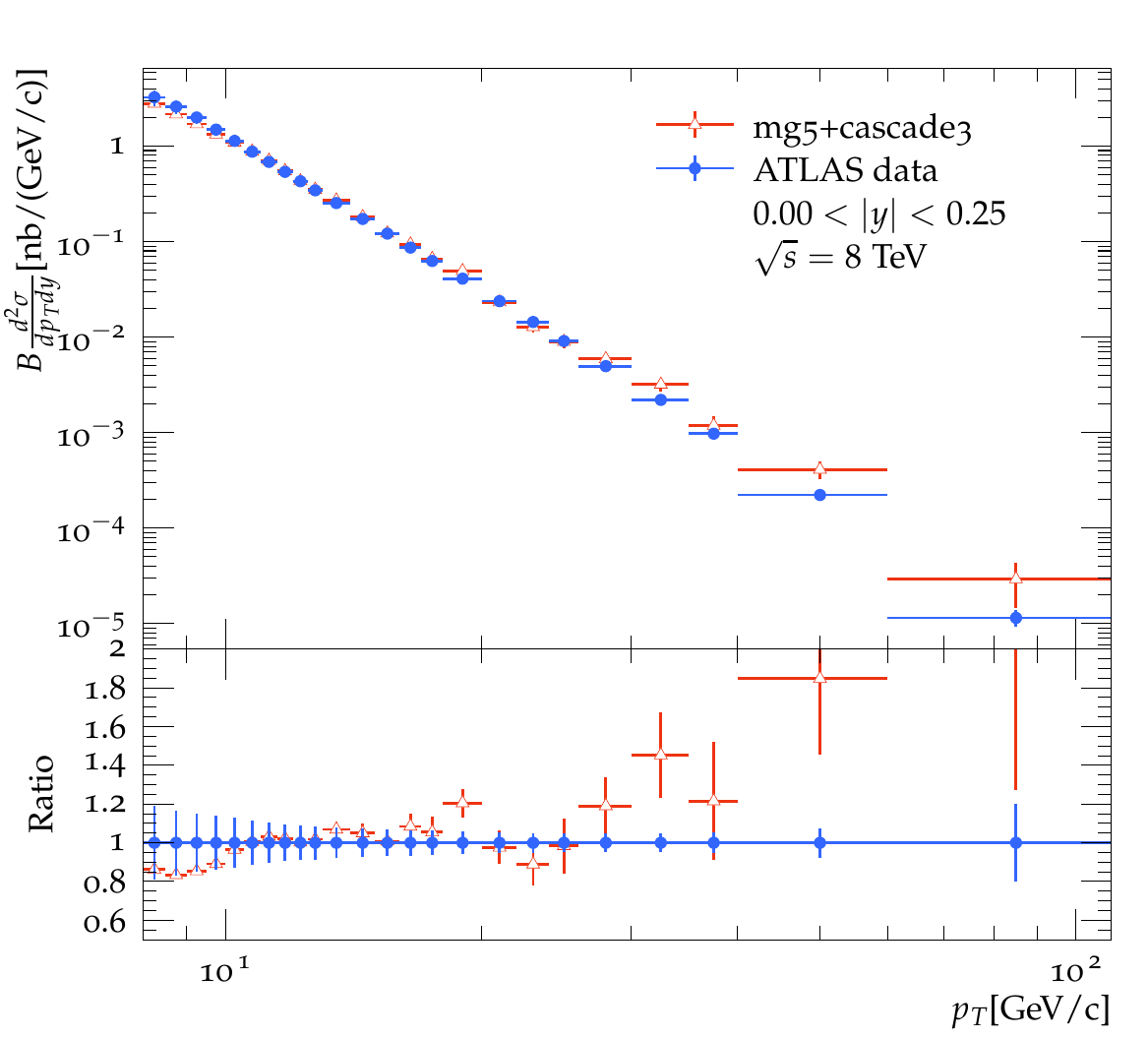}
\end{center}
\end{figure}

\begin{figure}[h!]
\begin{center}
 \includegraphics[width=22pc]{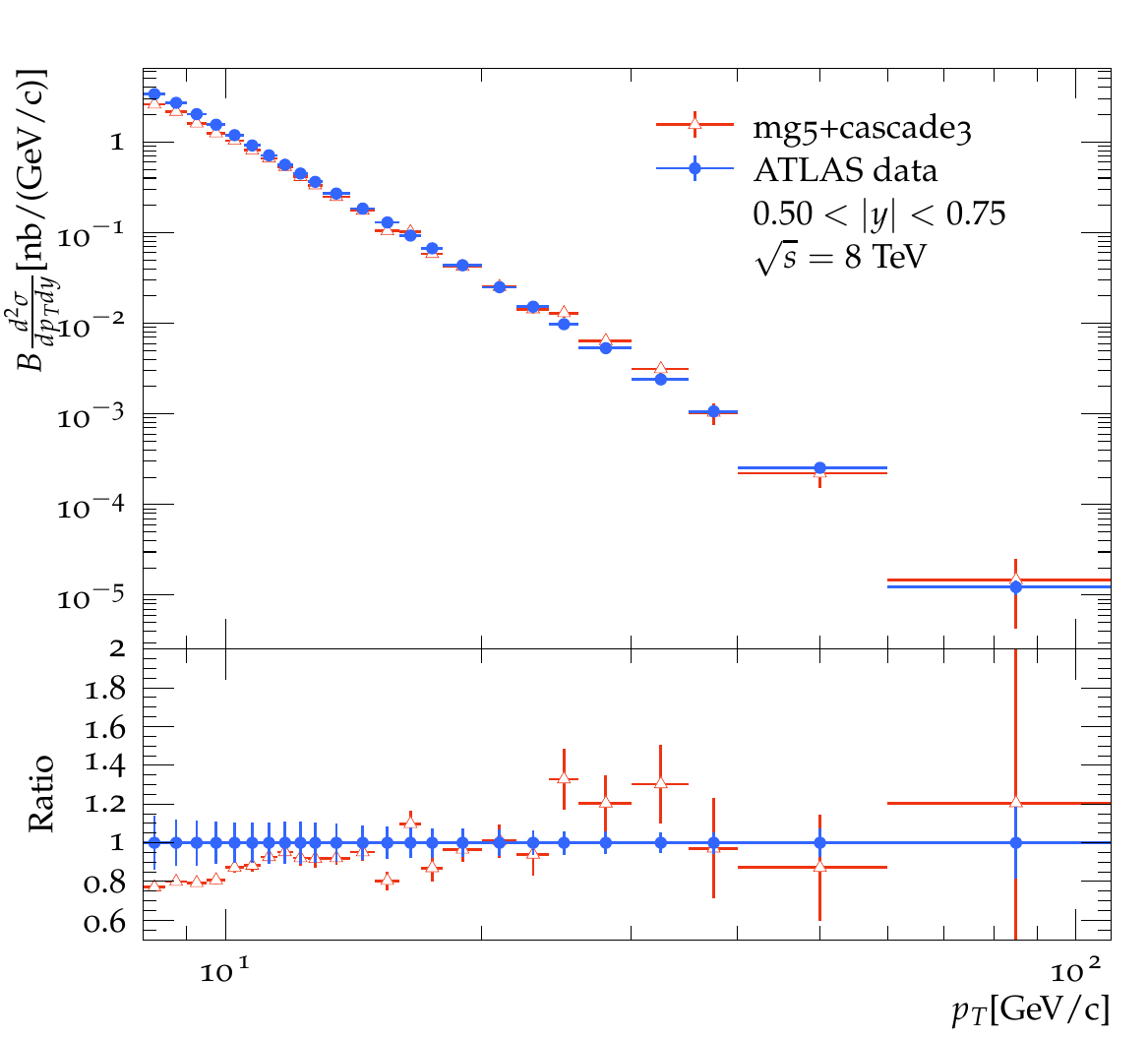}
\end{center}
\end{figure}

\begin{figure}[h!]
\begin{center}
 \includegraphics[width=22pc]{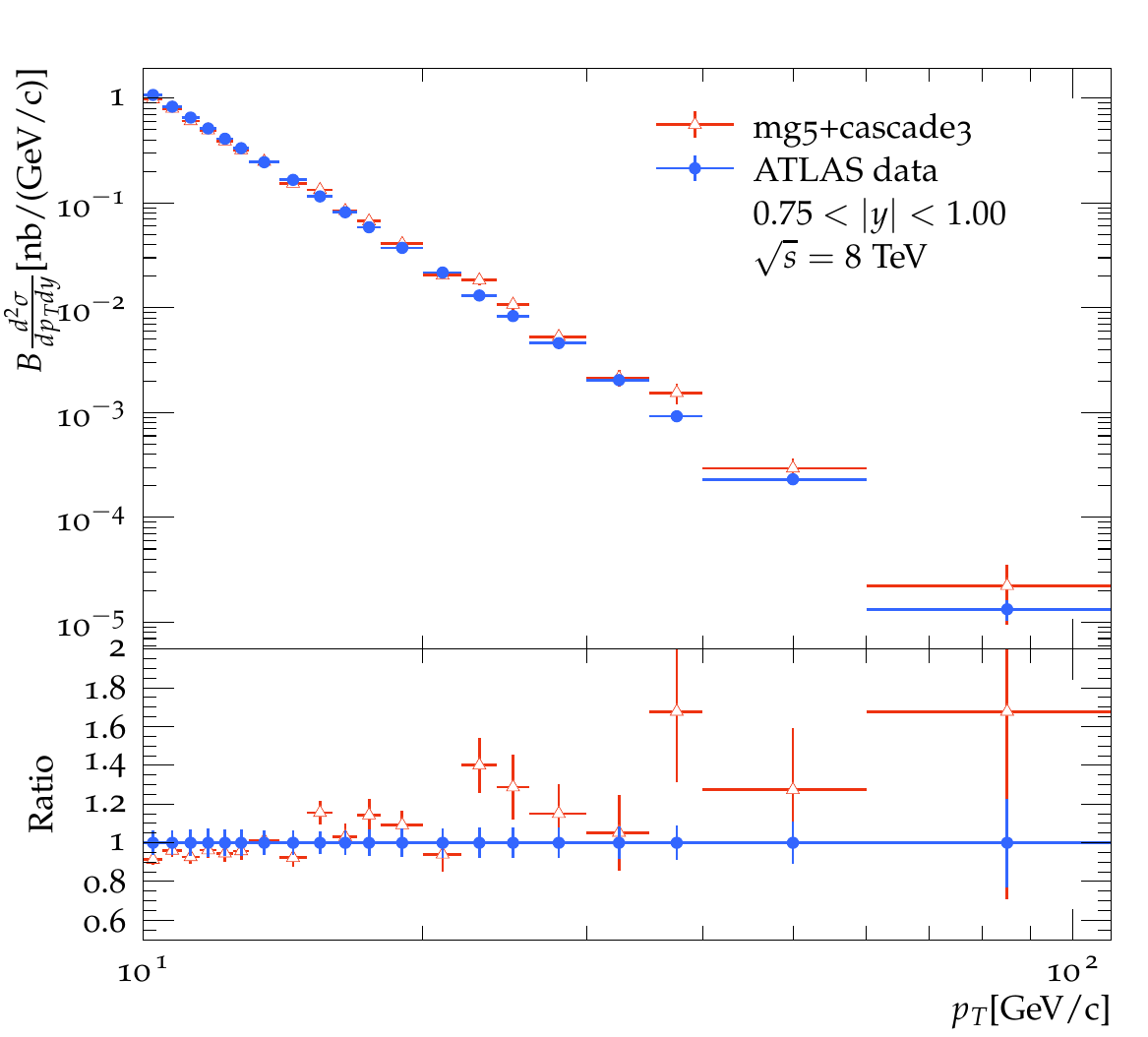}
\end{center}
\end{figure}

\begin{figure}[h!]
\begin{center}
 \includegraphics[width=22pc]{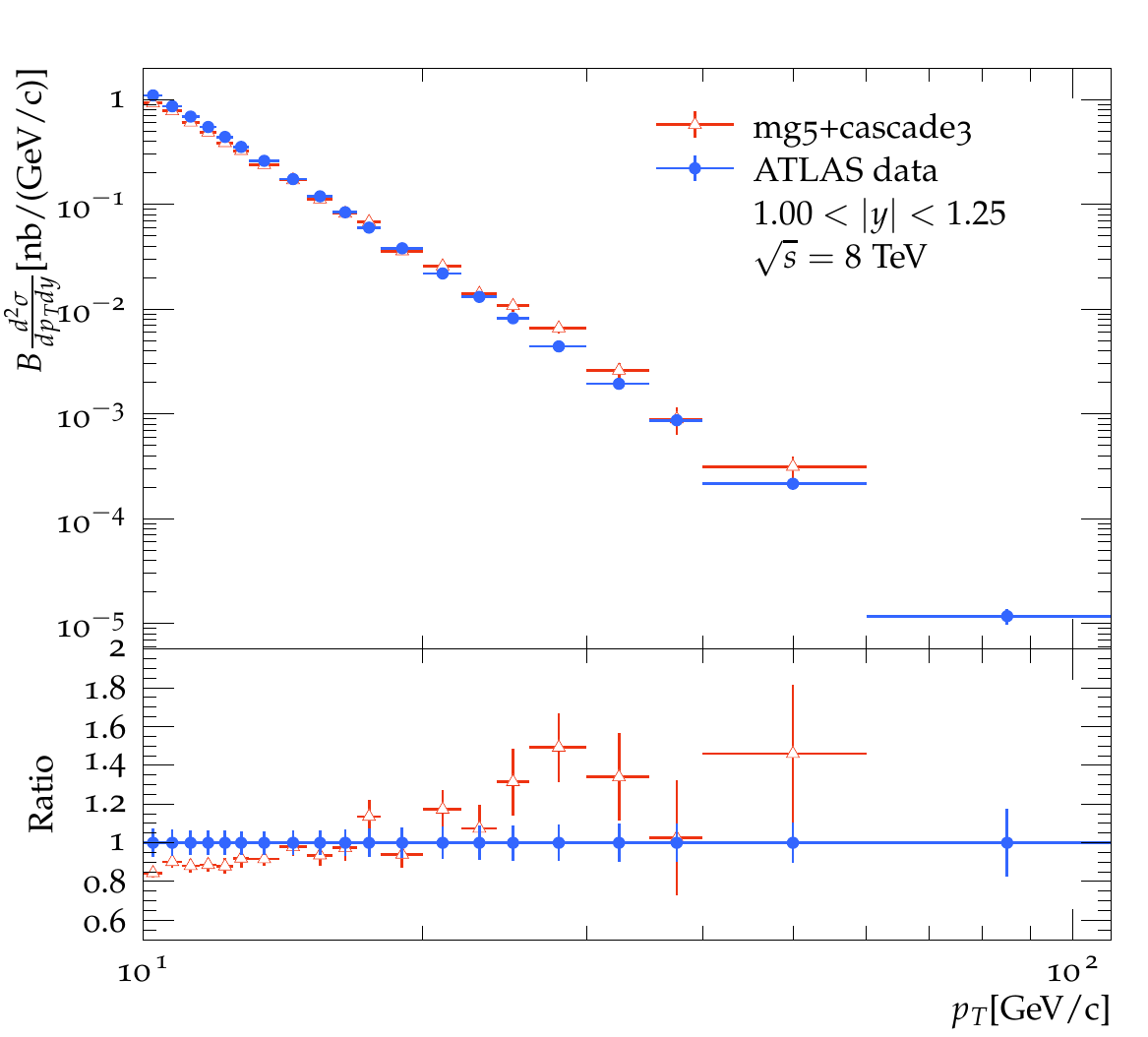}
\end{center}
\end{figure}

\begin{figure}[h!]
\begin{center}
 \includegraphics[width=22pc]{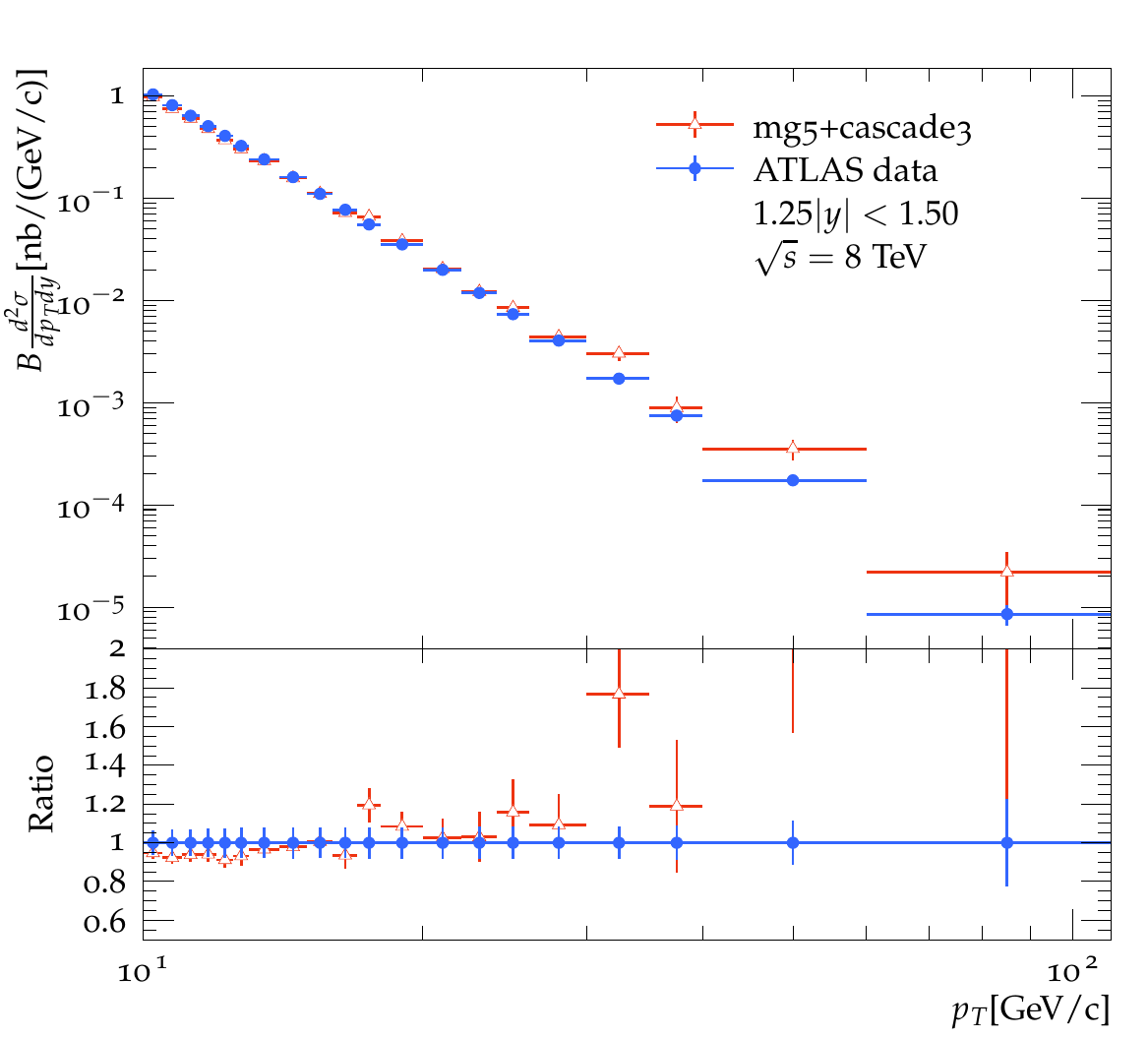}
\end{center}
\end{figure}

\begin{figure}[h!]
\begin{center}
 \includegraphics[width=22pc]{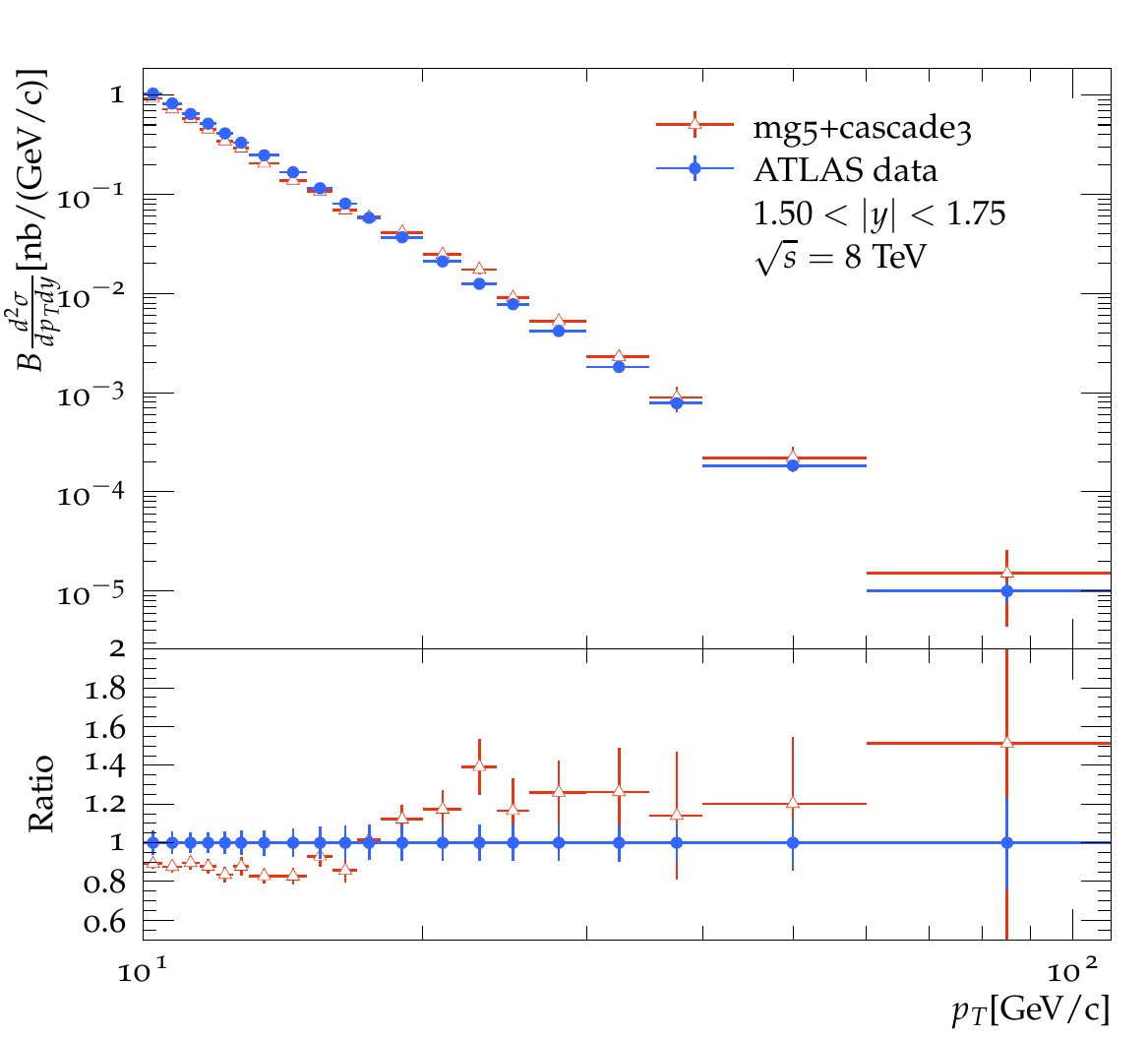}
\end{center}
\end{figure}

\begin{figure}[h!]
\begin{center}
 \includegraphics[width=22pc]{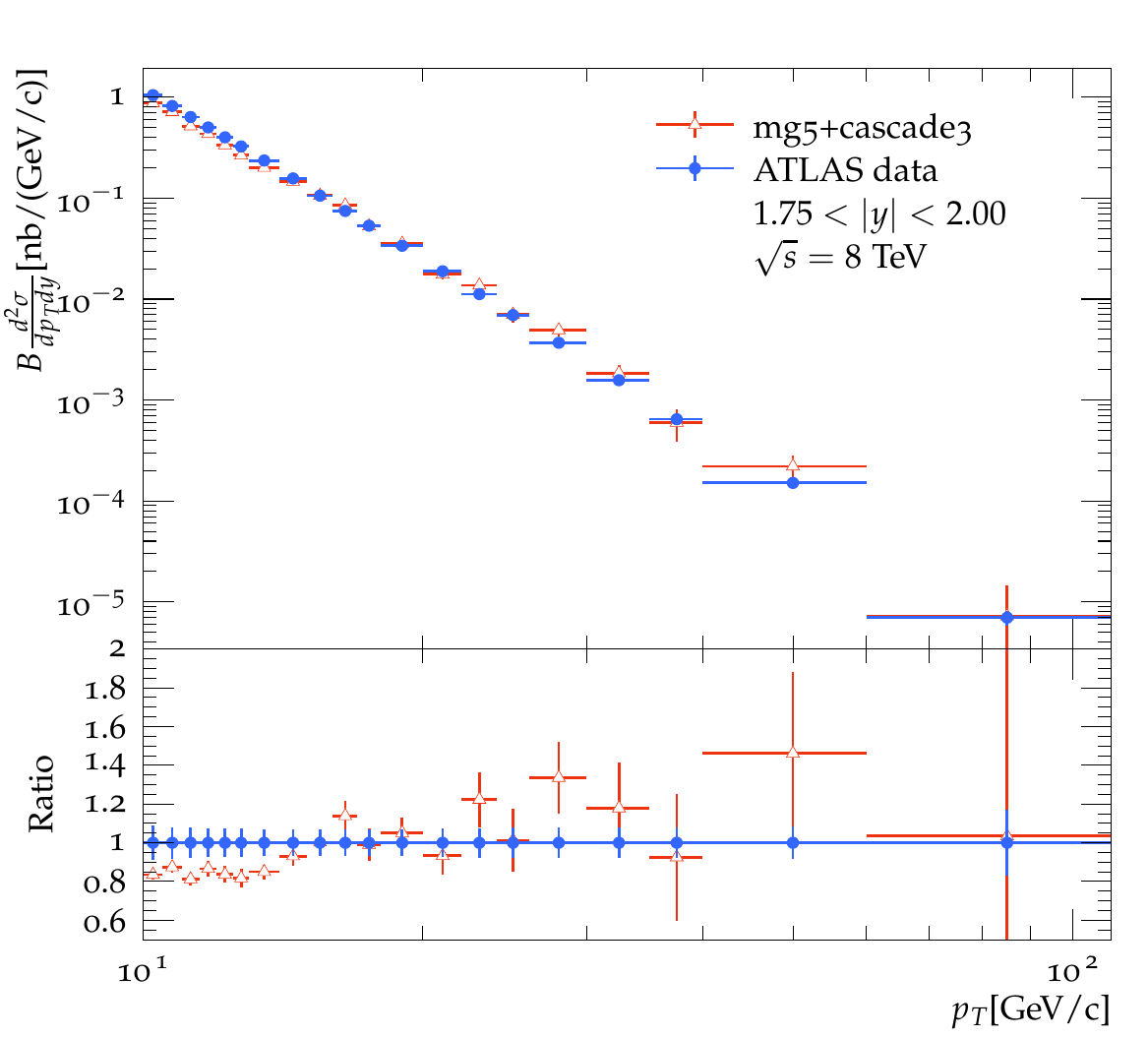}
\end{center}
\end{figure}

\clearpage

\section{Conclusion}
We have shown that the evolution of the $Q\bar{Q}$ pair predicted by pQCD, and in particular real emissions, solves the issue of NLO calculations with the CEM. However, one should keep in mind that our calculations used several approximations discussed in Sec.~\ref{secRes}. Still, there is a little doubt on the significant effect of the evolution on the quarkonium transverse-momentum distribution for factorization scales roughly larger than 20 GeV. Of course, this is also true for the usual fragmentation functions obeying to the DGLAP equation. We expect that the evolution could also have a significant impact on the determination of NRQCD LDMEs, for instance, at large $p_t$ or large $Q^2=-q^2$ in semi-inclusive DIS, where $q$ is the virtual-photon momentum. The latter case is relevant since the EIC will reach $Q$ lager than 20 GeV \cite{Khalek2022}.

In Sec.~\ref{secUni}, we showed that with the value $F_{J/\psi}=0.014$, we could describe data from $\sqrt{s}=13$ TeV to $\sqrt{s}=200$ GeV with satisfying accuracy. In other words, $F_{J/\psi}$ seems to be universal, as expected. Other calculations based on $k_t$-factorization  did not reach this conclusion, and we speculated on the possible explanations. One of these explanations involves the definition of UPDFs, and we believe that quarkonium data have the potential to constrain these functions.

\section*{Acknowledgments}
B.G. and I.S. acknowledge support from ANID PIA/APOYO AFB220004. I.S. is supported by Fondecyt project 1230391. We would like to thank H. Jung for his help with CASCADE3. We also thank H. Jung and F. Hautmann for interesting discussions.

\bibliographystyle{BibFiles/t1}
\bibliography{BibFiles/quarkoniaBib,BibFiles/GeneralBib}

\begin{thebibliography}{10}
\providecommand{\url}[1]{\texttt{#1}}
\providecommand{\urlprefix}{URL }
\providecommand{\eprint}[2][]{\url{#2}}

\bibitem{Bodwin1995}
G.~T. Bodwin, E.~Braaten and G.~P. Lepage, \emph{Rigorous {QCD} analysis of
  inclusive annihilation and production of heavy quarkonium},
  \MYhref[journalLinks]{http://dx.doi.org/10.1103/physrevd.51.1125}{Physical
  Review D
  }\MYhref[journalLinks]{http://dx.doi.org/10.1103/physrevd.51.1125}{\textbf{51}
  (1995) 3 1125--1171}.

\bibitem{Bodwin2005}
G.~T. Bodwin, E.~Braaten and J.~Lee, \emph{Comparison of the color-evaporation
  model and the nonrelativistic {QCD} factorization approach in charmonium
  production},
  \MYhref[journalLinks]{http://dx.doi.org/10.1103/physrevd.72.014004}{Physical
  Review D
  }\MYhref[journalLinks]{http://dx.doi.org/10.1103/physrevd.72.014004}{\textbf{72}
  (2005) 1 014004}.

\bibitem{Lansberg2020}
J.-P. Lansberg, \emph{New observables in inclusive production of quarkonia},
  \MYhref[journalLinks]{http://dx.doi.org/10.1016/j.physrep.2020.08.007}{Physics
  Reports
  }\MYhref[journalLinks]{http://dx.doi.org/10.1016/j.physrep.2020.08.007}{\textbf{889}
  (2020) 1--106}.

\bibitem{Alwall2014}
J.~Alwall et~al., \emph{The automated computation of tree-level and
  next-to-leading order differential cross sections, and their matching to
  parton shower simulations},
  \MYhref[journalLinks]{http://dx.doi.org/10.1007/jhep07(2014)079}{JHEP
  }\MYhref[journalLinks]{http://dx.doi.org/10.1007/jhep07(2014)079}{\textbf{07}
  (2014) 079}.

\bibitem{Aad2016}
G.~Aad et~al., \emph{Measurement of the differential cross-sections of prompt
  and non-prompt production of {J}$/\psi$ and $\psi (2s)$ in pp collisions at
  $\sqrt{s}=7$ and $8$ {T}e{V} with the {ATLAS} detector},
  \MYhref[journalLinks]{http://dx.doi.org/10.1140/epjc/s10052-016-4050-8}{The
  European Physical Journal C
  }\MYhref[journalLinks]{http://dx.doi.org/10.1140/epjc/s10052-016-4050-8}{\textbf{76}
  (2016) 5}.

\bibitem{Kang2011}
Z.-B. Kang, J.-W. Qiu and G.~Sterman, \emph{Factorization and {Q}uarkonium
  {P}roduction},
  \MYhref[journalLinks]{http://dx.doi.org/10.1016/j.nuclphysbps.2011.03.054}{Nuclear
  Physics B - Proceedings Supplements
  }\MYhref[journalLinks]{http://dx.doi.org/10.1016/j.nuclphysbps.2011.03.054}{\textbf{214}
  (2011) 1 39--43}.

\bibitem{Kang2012}
Z.-B. Kang, J.-W. Qiu and G.~Sterman, \emph{Heavy {Q}uarkonium {P}roduction and
  {P}olarization},
  \MYhref[journalLinks]{http://dx.doi.org/10.1103/physrevlett.108.102002}{Physical
  Review Letters
  }\MYhref[journalLinks]{http://dx.doi.org/10.1103/physrevlett.108.102002}{\textbf{108}
  (2012) 10 102002}.

\bibitem{Kang2014}
Z.-B. Kang, Y.-Q. Ma, J.-W. Qiu and G.~Sterman, \emph{Heavy quarkonium
  production at collider energies: Factorization and evolution},
  \MYhref[journalLinks]{http://dx.doi.org/10.1103/physrevd.90.034006}{Physical
  Review D
  }\MYhref[journalLinks]{http://dx.doi.org/10.1103/physrevd.90.034006}{\textbf{90}
  (2014) 3 034006}.

\bibitem{Lee2022}
K.~Lee, J.-W. Qiu, G.~Sterman and K.~Watanabe, \emph{Subleading power
  corrections to heavy quarkonium production in {QCD} factorization approach},
  \MYhref[journalLinks]{http://dx.doi.org/10.1051/epjconf/202227404005}{{EPJ}
  Web of Conferences
  }\MYhref[journalLinks]{http://dx.doi.org/10.1051/epjconf/202227404005}{\textbf{274}
  (2022) 04005}.

\bibitem{Fleming2012}
S.~Fleming, A.~K. Leibovich, T.~Mehen and I.~Z. Rothstein, \emph{Systematics of
  quarkonium production at the {LHC} and double parton fragmentation},
  \MYhref[journalLinks]{http://dx.doi.org/10.1103/physrevd.86.094012}{Physical
  Review D
  }\MYhref[journalLinks]{http://dx.doi.org/10.1103/physrevd.86.094012}{\textbf{86}
  (2012) 9 094012}.

\bibitem{Fleming2013}
S.~Fleming, A.~K. Leibovich, T.~Mehen and I.~Z. Rothstein, \emph{Anomalous
  dimensions of the double parton fragmentation functions},
  \MYhref[journalLinks]{http://dx.doi.org/10.1103/physrevd.87.074022}{Physical
  Review D
  }\MYhref[journalLinks]{http://dx.doi.org/10.1103/physrevd.87.074022}{\textbf{87}
  (2013) 7 074022}.

\bibitem{Fritzsch1977}
H.~Fritzsch, \emph{Producing heavy quark flavors in hadronic
  collisions{\textemdash}{\textquotesingle} {A} test of quantum
  chromodynamics},
  \MYhref[journalLinks]{http://dx.doi.org/10.1016/0370-2693(77)90108-3}{Physics
  Letters B
  }\MYhref[journalLinks]{http://dx.doi.org/10.1016/0370-2693(77)90108-3}{\textbf{67}
  (1977) 2 217--221}.

\bibitem{Halzen1977}
F.~Halzen, \emph{{CVC} for gluons and hadroproduction of quark flavours},
  \MYhref[journalLinks]{http://dx.doi.org/10.1016/0370-2693(77)90144-7}{Physics
  Letters B
  }\MYhref[journalLinks]{http://dx.doi.org/10.1016/0370-2693(77)90144-7}{\textbf{69}
  (1977) 1 105--108}.

\bibitem{Sjoestrand2006}
T.~Sj\"ostrand, S.~Mrenna and P.~Skands, \emph{{PYTHIA} 6.4 physics and
  manual},
  \MYhref[journalLinks]{http://dx.doi.org/10.1088/1126-6708/2006/05/026}{JHEP
  }\MYhref[journalLinks]{http://dx.doi.org/10.1088/1126-6708/2006/05/026}{\textbf{05}
  (2006) 026}.

\bibitem{GribovLipatov:1972}
V.~N. Gribov and L.~N. Lipatov, \emph{Deep inelastic e p scattering in
  perturbation theory}, Sov. J. Nucl. Phys. \textbf{15} (1972) 438.

\bibitem{AltarelliParisi:1977}
G.~Altarelli and G.~Parisi, \emph{Asymptotic freedom in parton language},
  \MYhref[journalLinks]{http://dx.doi.org/10.1016/0550-3213(77)90384-4}{Nuclear
  Physics B
  }\MYhref[journalLinks]{http://dx.doi.org/10.1016/0550-3213(77)90384-4}{\textbf{126}
  (1977) 298--318}.

\bibitem{Dokshitzer:1977}
Y.~L. Dokshitzer, \emph{Calculation of the {S}tructure {F}unctions for {D}eep
  {I}nelastic {S}cattering and e+ e- {A}nnihilation by {P}erturbation {T}heory
  in {Q}uantum {C}hromodynamics.}, Sov. Phys. JETP \textbf{46} (1977) 641.

\bibitem{Ellis1996}
R.~K. Ellis, W.~J. Stirling and B.~R. Webber, \emph{{QCD} and Collider
  Physics}, Cambridge University Press (1996).

\bibitem{Echevarria2019}
M.~G. Echevarria, \emph{Proper {TMD} factorization for quarkonia production:
  $pp \to \eta_{c,b}$ as a study case},
  \MYhref[journalLinks]{http://dx.doi.org/10.1007/jhep10(2019)144}{JHEP
  }\MYhref[journalLinks]{http://dx.doi.org/10.1007/jhep10(2019)144}{\textbf{10}
  (2019) 144}.

\bibitem{Grlery}
L.~Gribov, E.~Levin and M.~Ryskin, \emph{Semihard processes in {QCD}}, Physics
  Reports \textbf{100} (1983) 1.

\bibitem{Catani1990}
S.~Catani, M.~Ciafaloni and F.~Hautmann, \emph{Gluon contributions to small $x$
  heavy flavour production},
  \MYhref[journalLinks]{http://dx.doi.org/10.1016/0370-2693(90)91601-7}{Physics
  Letters B
  }\MYhref[journalLinks]{http://dx.doi.org/10.1016/0370-2693(90)91601-7}{\textbf{242}
  (1990) 1 97--102}.

\bibitem{Collins1991}
J.~Collins and R.~Ellis, \emph{Heavy-quark production in very high energy
  hadron collisions},
  \MYhref[journalLinks]{http://dx.doi.org/10.1016/0550-3213(91)90288-9}{Nuclear
  Physics B
  }\MYhref[journalLinks]{http://dx.doi.org/10.1016/0550-3213(91)90288-9}{\textbf{360}
  (1991) 1 3--30}.

\bibitem{Catani1991}
S.~Catani, M.~Ciafaloni and F.~Hautmann, \emph{High energy factorization and
  small-x heavy flavour production},
  \MYhref[journalLinks]{http://dx.doi.org/10.1016/0550-3213(91)90055-3}{Nuclear
  Physics B
  }\MYhref[journalLinks]{http://dx.doi.org/10.1016/0550-3213(91)90055-3}{\textbf{366}
  (1991) 1 135--188}.

\bibitem{Lerysh}
E.~M. Levin, M.~G. Ryskin, Y.~M. Shabelski and A.~G. Shuvaev, \emph{Heavy quark
  production in semihard nucleon interaction}, Sov. J. Nucl. Phys. \textbf{53}
  (1991) 657.

\bibitem{Baranov2002}
S.~P. Baranov, \emph{Highlights from the $k_t$-factorization approach on the
  quarkonium production puzzles},
  \MYhref[journalLinks]{http://dx.doi.org/10.1103/physrevd.66.114003}{Physical
  Review D
  }\MYhref[journalLinks]{http://dx.doi.org/10.1103/physrevd.66.114003}{\textbf{66}
  (2002) 11 114003}.

\bibitem{Jung2011}
H.~Jung, M.~Kraemer, A.~V. Lipatov and N.~P. Zotov, \emph{Beauty quark and
  quarkonium production at {LHC}: kt-factorization and {CASCADE} versus data}
  (2011), \MYhref[eprintLinks]{http://arxiv.org/abs/1107.4328}{{\ttfamily
  arXiv:1107.4328 [hep-ph]}}.

\bibitem{Saleev2012}
V.~A. Saleev, M.~A. Nefedov and A.~V. Shipilova, \emph{Prompt {J}$/\psi$
  production in the regge limit of {QCD}: {F}rom the tevatron to the {LHC}},
  \MYhref[journalLinks]{http://dx.doi.org/10.1103/physrevd.85.074013}{Physical
  Review D
  }\MYhref[journalLinks]{http://dx.doi.org/10.1103/physrevd.85.074013}{\textbf{85}
  (2012) 7 074013}.

\bibitem{Baranov2012}
S.~P. Baranov, A.~V. Lipatov and N.~P. Zotov, \emph{Prompt {J}$/\psi$
  production at {LHC}: new evidence for the $k_t$-factorization},
  \MYhref[journalLinks]{http://dx.doi.org/10.1103/physrevd.85.014034}{Physical
  Review D
  }\MYhref[journalLinks]{http://dx.doi.org/10.1103/physrevd.85.014034}{\textbf{85}
  (2012) 1 014034}.

\bibitem{Cheung2018}
V.~Cheung and R.~Vogt, \emph{Production and polarization of prompt {J}$/\psi$
  in the improved color evaporation model using the $k_t$-factorization
  approach},
  \MYhref[journalLinks]{http://dx.doi.org/10.1103/physrevd.98.114029}{Physical
  Review D
  }\MYhref[journalLinks]{http://dx.doi.org/10.1103/physrevd.98.114029}{\textbf{98}
  (2018) 11 114029}.

\bibitem{Cisek2018}
A.~Cisek and A.~Szczurek, \emph{Prompt inclusive production of {J}$/\psi$,
  $\psi'$ and $\chi_c$ mesons at the {LHC} in forward directions within the
  {NRQCD} $k_t$-factorization approach: {S}earch for the onset of gluon
  saturation},
  \MYhref[journalLinks]{http://dx.doi.org/10.1103/physrevd.97.034035}{Physical
  Review D
  }\MYhref[journalLinks]{http://dx.doi.org/10.1103/physrevd.97.034035}{\textbf{97}
  (2018) 3 034035}.

\bibitem{Maciula2019}
R.~Maciu{\l}a, A.~Szczurek and A.~Cisek, \emph{J$/\psi$-meson production within
  improved color evaporation model with the $k_t$-factorization approach for
  $c\bar{c}$ production},
  \MYhref[journalLinks]{http://dx.doi.org/10.1103/physrevd.99.054014}{Physical
  Review D
  }\MYhref[journalLinks]{http://dx.doi.org/10.1103/physrevd.99.054014}{\textbf{99}
  (2019) 5 054014}.

\bibitem{Baranov2019}
S.~Baranov and A.~Lipatov, \emph{Are there any challenges in the charmonia
  production and polarization at the {LHC}?},
  \MYhref[journalLinks]{http://dx.doi.org/10.1103/physrevd.100.114021}{Physical
  Review D
  }\MYhref[journalLinks]{http://dx.doi.org/10.1103/physrevd.100.114021}{\textbf{100}
  (2019) 11 114021}.

\bibitem{Chernyshev2022}
A.~Chernyshev and V.~Saleev, \emph{Single and pair {J}$/\psi$ production in the
  improved color evaporation model using the parton reggeization approach},
  \MYhref[journalLinks]{http://dx.doi.org/10.1103/physrevd.106.114006}{Physical
  Review D
  }\MYhref[journalLinks]{http://dx.doi.org/10.1103/physrevd.106.114006}{\textbf{106}
  (2022) 11 114006}.

\bibitem{Martinez2020}
A.~B. Martinez et~al., \emph{The transverse momentum spectrum of low mass
  {D}rell-{Y}an production at next-to-leading order in the parton branching
  method},
  \MYhref[journalLinks]{http://dx.doi.org/10.1140/epjc/s10052-020-8136-y}{Eur.
  Phys. J. C
  }\MYhref[journalLinks]{http://dx.doi.org/10.1140/epjc/s10052-020-8136-y}{\textbf{80}
  (2020) 7}.

\bibitem{Martinez2019}
A.~B. Martinez et~al., \emph{Collinear and {TMD} parton densities from fits to
  precision {DIS} measurements in the parton branching method},
  \MYhref[journalLinks]{http://dx.doi.org/10.1103/physrevd.99.074008}{Physical
  Review D
  }\MYhref[journalLinks]{http://dx.doi.org/10.1103/physrevd.99.074008}{\textbf{99}
  (2019) 7 074008}.

\bibitem{Hautmann2017}
F.~Hautmann et~al., \emph{Soft-gluon resolution scale in {QCD} evolution
  equations},
  \MYhref[journalLinks]{http://dx.doi.org/10.1016/j.physletb.2017.07.005}{Physics
  Letters B
  }\MYhref[journalLinks]{http://dx.doi.org/10.1016/j.physletb.2017.07.005}{\textbf{772}
  (2017) 446--451}.

\bibitem{Hautmann2018}
F.~Hautmann et~al., \emph{Collinear and {TMD} quark and gluon densities from
  parton branching solution of {QCD} evolution equations},
  \MYhref[journalLinks]{http://dx.doi.org/10.1007/jhep01(2018)070}{JHEP
  }\MYhref[journalLinks]{http://dx.doi.org/10.1007/jhep01(2018)070}{\textbf{01}
  (2018) 070}.

\bibitem{Martinez2019a}
A.~B. Martinez et~al., \emph{Production of ${Z}$-bosons in the parton branching
  method},
  \MYhref[journalLinks]{http://dx.doi.org/10.1103/physrevd.100.074027}{Physical
  Review D
  }\MYhref[journalLinks]{http://dx.doi.org/10.1103/physrevd.100.074027}{\textbf{100}
  (2019) 7 074027}.

\bibitem{Martinez2021}
A.~B. Martinez, F.~Hautmann and M.~Mangano, \emph{{TMD} evolution and multi-jet
  merging},
  \MYhref[journalLinks]{http://dx.doi.org/10.1016/j.physletb.2021.136700}{Physics
  Letters B
  }\MYhref[journalLinks]{http://dx.doi.org/10.1016/j.physletb.2021.136700}{\textbf{822}
  (2021) 136700}.

\bibitem{Martinez2022}
A.~B. Martinez, F.~Hautmann and M.~L. Mangano, \emph{Multi-jet merging with
  {TMD} parton branching},
  \MYhref[journalLinks]{http://dx.doi.org/10.1007/jhep09(2022)060}{JHEP
  }\MYhref[journalLinks]{http://dx.doi.org/10.1007/jhep09(2022)060}{\textbf{09}
  (2022) 060s}.

\bibitem{Jung2010}
H.~Jung et~al., \emph{The {CCFM} monte carlo generator {CASCADE}
  version~2.2.03},
  \MYhref[journalLinks]{http://dx.doi.org/10.1140/epjc/s10052-010-1507-z}{The
  European Physical Journal C
  }\MYhref[journalLinks]{http://dx.doi.org/10.1140/epjc/s10052-010-1507-z}{\textbf{70}
  (2010) 4 1237--1249}.

\bibitem{Baranov2021}
S.~Baranov et~al., \emph{{CASCADE}3 a monte carlo event generator based on
  {TMDs}},
  \MYhref[journalLinks]{http://dx.doi.org/10.1140/epjc/s10052-021-09203-8}{The
  European Physical Journal C
  }\MYhref[journalLinks]{http://dx.doi.org/10.1140/epjc/s10052-021-09203-8}{\textbf{81}
  (2021) 5}.

\bibitem{Dulat2016}
S.~Dulat et~al., \emph{New parton distribution functions from a global analysis
  of quantum chromodynamics},
  \MYhref[journalLinks]{http://dx.doi.org/10.1103/physrevd.93.033006}{Physical
  Review D
  }\MYhref[journalLinks]{http://dx.doi.org/10.1103/physrevd.93.033006}{\textbf{93}
  (2016) 3 033006}.

\bibitem{Thorne2008}
R.~S. Thorne and W.~K. Tung, \emph{P{QCD} formulations with {H}eavy {Q}uark
  {M}asses and {G}lobal {A}nalysis}  (2008),
  \MYhref[eprintLinks]{http://arxiv.org/abs/0809.0714}{{\ttfamily
  arXiv:0809.0714 [hep-ph]}}.

\bibitem{Ma2016}
Y.-Q. Ma and R.~Vogt, \emph{Quarkonium production in an improved color
  evaporation model},
  \MYhref[journalLinks]{http://dx.doi.org/10.1103/physrevd.94.114029}{Physical
  Review D
  }\MYhref[journalLinks]{http://dx.doi.org/10.1103/physrevd.94.114029}{\textbf{94}
  (2016) 11 114029}.

\bibitem{Lansberg2020a}
J.-P. Lansberg et~al., \emph{Complete {NLO} {QCD} study of single- and
  double-quarkonium hadroproduction in the colour-evaporation model at the
  tevatron and the {LHC}},
  \MYhref[journalLinks]{http://dx.doi.org/10.1016/j.physletb.2020.135559}{Physics
  Letters B
  }\MYhref[journalLinks]{http://dx.doi.org/10.1016/j.physletb.2020.135559}{\textbf{807}
  (2020) 135559}.

\bibitem{Kang2015}
Z.-B. Kang, Y.-Q. Ma, J.-W. Qiu and G.~Sterman, \emph{Heavy quarkonium
  production at collider energies: {P}artonic cross section and polarization},
  \MYhref[journalLinks]{http://dx.doi.org/10.1103/physrevd.91.014030}{Physical
  Review D
  }\MYhref[journalLinks]{http://dx.doi.org/10.1103/physrevd.91.014030}{\textbf{91}
  (2015) 1 014030}.

\bibitem{LHCbQ2011}
R.~Aaij et~al., \emph{Measurement of {J}$/\psi$ production in pp collisions at
  $\sqrt{s}$=7 {T}e{V}},
  \MYhref[journalLinks]{http://dx.doi.org/10.1140/epjc/s10052-011-1645-y}{Eur.
  Phys. J. C
  }\MYhref[journalLinks]{http://dx.doi.org/10.1140/epjc/s10052-011-1645-y}{\textbf{71}
  (2011) 1645}, \MYhref[eprintLinks]{http://arxiv.org/abs/1103.0423}{{\ttfamily
  arXiv:1103.0423 [hep-ex]}}.

\bibitem{AliceQ2012}
B.~Abelev et~al., \emph{Measurement of prompt {J}/$\psi$ and beauty hadron
  production cross sections at mid-rapidity in pp collisions at $\sqrt{s}=7$
  {T}e{V}},
  \MYhref[journalLinks]{http://dx.doi.org/10.1007/jhep11(2012)065}{JHEP
  }\MYhref[journalLinks]{http://dx.doi.org/10.1007/jhep11(2012)065}{\textbf{11}
  (2012) 065}.

\bibitem{Chatrchyan2012}
S.~Chatrchyan et~al., \emph{J/$\psi$ and $\psi$(2s) production in pp collisions
  at $\sqrt{s} = 7$ {T}e{V}},
  \MYhref[journalLinks]{http://dx.doi.org/10.1007/jhep02(2012)011}{JHEP
  }\MYhref[journalLinks]{http://dx.doi.org/10.1007/jhep02(2012)011}{\textbf{02}
  (2012) 011}.

\bibitem{Bain2017}
R.~Bain et~al., \emph{{NRQCD} confronts {LHCb} data on quarkonium production
  within jets},
  \MYhref[journalLinks]{http://dx.doi.org/10.1103/physrevlett.119.032002}{Physical
  Review Letters
  }\MYhref[journalLinks]{http://dx.doi.org/10.1103/physrevlett.119.032002}{\textbf{119}
  (2017) 3 032002}.

\bibitem{Baumgart2014}
M.~Baumgart, A.~K. Leibovich, T.~Mehen and I.~Z. Rothstein, \emph{Probing
  quarkonium production mechanisms with jet substructure},
  \MYhref[journalLinks]{http://dx.doi.org/10.1007/jhep11(2014)003}{JHEP
  }\MYhref[journalLinks]{http://dx.doi.org/10.1007/jhep11(2014)003}{\textbf{11}
  (2014) 003}.

\bibitem{Aaij2015}
R.~Aaij et~al., \emph{Measurement of forward {J}/$\psi$ production
  cross-sections in pp collisions at $\sqrt{s} = 13$ {T}e{V}},
  \MYhref[journalLinks]{http://dx.doi.org/10.1007/jhep10(2015)172}{JHEP
  }\MYhref[journalLinks]{http://dx.doi.org/10.1007/jhep10(2015)172}{\textbf{10}
  (2015) 172}.

\bibitem{Aaij2017}
R.~Aaij et~al., \emph{Erratum to: Measurement of forward {J}/$\psi$ production
  cross-sections in pp collisions at $\sqrt{s} = 13$ {T}e{V}},
  \MYhref[journalLinks]{http://dx.doi.org/10.1007/jhep05(2017)063}{JHEP
  }\MYhref[journalLinks]{http://dx.doi.org/10.1007/jhep05(2017)063}{\textbf{05}
  (2017) 063}.

\bibitem{Adare2007}
A.~Adare et~al., \emph{J/$\psi$ {P}roduction versus {T}ransverse {M}omentum and
  {R}apidity in $p+p$ collisions at $\sqrt{s}=200$ {G}e{V}},
  \MYhref[journalLinks]{http://dx.doi.org/10.1103/physrevlett.98.232002}{Physical
  Review Letters
  }\MYhref[journalLinks]{http://dx.doi.org/10.1103/physrevlett.98.232002}{\textbf{98}
  (2007) 23 232002}.

\bibitem{Kimber2001}
M.~A. Kimber, A.~D. Martin and M.~G. Ryskin, \emph{Unintegrated parton
  distributions},
  \MYhref[journalLinks]{http://dx.doi.org/10.1103/physrevd.63.114027}{Physical
  Review D
  }\MYhref[journalLinks]{http://dx.doi.org/10.1103/physrevd.63.114027}{\textbf{63}
  (2001) 11 114027}.

\bibitem{Watt2003}
G.~Watt, A.~D. Martin and M.~G. Ryskin, \emph{Unintegrated parton distributions
  and inclusive jet productionat {HERA}},
  \MYhref[journalLinks]{http://dx.doi.org/10.1140/epjc/s2003-01320-4}{The
  European Physical Journal C
  }\MYhref[journalLinks]{http://dx.doi.org/10.1140/epjc/s2003-01320-4}{\textbf{31}
  (2003) 1 73--89}.

\bibitem{Guiot2020a}
B.~Guiot, \emph{{P}athologies of the {K}imber-{M}artin-{R}yskin prescriptions
  for unintegrated {PDFs}: Which prescription should be preferred?},
  \MYhref[journalLinks]{http://dx.doi.org/10.1103/physrevd.101.054006}{Physical
  Review D
  }\MYhref[journalLinks]{http://dx.doi.org/10.1103/physrevd.101.054006}{\textbf{101}
  (2020) 5 054006}.

\bibitem{Guiot2021}
B.~Guiot and A.~van Hameren, \emph{${D}$ and ${B}$-meson production using
  $k_t$-factorization calculations in a variable-flavor-number scheme},
  \MYhref[journalLinks]{http://dx.doi.org/10.1103/physrevd.104.094038}{Physical
  Review D
  }\MYhref[journalLinks]{http://dx.doi.org/10.1103/physrevd.104.094038}{\textbf{104}
  (2021) 9 094038}.

\bibitem{Guiot2023}
B.~Guiot, \emph{Normalization of unintegrated parton densities},
  \MYhref[journalLinks]{http://dx.doi.org/10.1103/physrevd.107.014015}{Physical
  Review D
  }\MYhref[journalLinks]{http://dx.doi.org/10.1103/physrevd.107.014015}{\textbf{107}
  (2023) 1 014015}.

\bibitem{Guiot2019}
B.~Guiot, \emph{Heavy-quark production with $k_t$ factorization: {T}he
  importance of the sea-quark distribution},
  \MYhref[journalLinks]{http://dx.doi.org/10.1103/physrevd.99.074006}{Physical
  Review D
  }\MYhref[journalLinks]{http://dx.doi.org/10.1103/physrevd.99.074006}{\textbf{99}
  (2019) 7 074006}.

\bibitem{Hautmann2019}
F.~Hautmann, L.~Keersmaekers, A.~Lelek and A.~van Kampen, \emph{Dynamical
  resolution scale in transverse momentum distributions at the {LHC}},
  \MYhref[journalLinks]{http://dx.doi.org/10.1016/j.nuclphysb.2019.114795}{Nuclear
  Physics B
  }\MYhref[journalLinks]{http://dx.doi.org/10.1016/j.nuclphysb.2019.114795}{\textbf{949}
  (2019) 114795}.

\bibitem{Werner2023a}
K.~Werner and B.~Guiot, \emph{Perturbative {QCD} concerning light and heavy
  flavor in the {EPOS}4 framework},
  \MYhref[journalLinks]{http://dx.doi.org/10.1103/physrevc.108.034904}{Physical
  Review C
  }\MYhref[journalLinks]{http://dx.doi.org/10.1103/physrevc.108.034904}{\textbf{108}
  (2023) 3 034904}, ISSN 2469-9993.

\bibitem{Gribov:1967vfb}
V.~N. Gribov, \emph{{A REGGEON DIAGRAM TECHNIQUE}}, Zh. Eksp. Teor. Fiz.
  \textbf{53} (1967) 654.

\bibitem{Gribov:1968jf}
V.~N. Gribov, \emph{Glauber corrections and the interaction between high-energy
  hadrons and nuclei}, Sov. Phys. JETP \textbf{29} (1969) 483.

\bibitem{Abramovsky:1973fm}
V.~A. Abramovsky, V.~N. Gribov and O.~V. Kancheli, \emph{Character of
  {I}nclusive {S}pectra and {F}luctuations {P}roduced in {I}nelastic
  {P}rocesses by {M}ulti-{P}omeron {E}xchange}, Yad. Fiz. \textbf{18} (1973)
  595.

\bibitem{Werner2023}
K.~Werner, \emph{Revealing a deep connection between factorization and
  saturation: {N}ew insight into modeling high-energy proton-proton and
  nucleus-nucleus scattering in the {EPOS}4 framework},
  \MYhref[journalLinks]{http://dx.doi.org/10.1103/physrevc.108.064903}{Physical
  Review C
  }\MYhref[journalLinks]{http://dx.doi.org/10.1103/physrevc.108.064903}{\textbf{108}
  (2023) 6 064903}, ISSN 2469-9993.

\bibitem{Khalek2022}
R.~A. Khalek et~al., \emph{Science {R}equirements and {D}etector {C}oncepts for
  the {E}lectron-{I}on {C}ollider},
  \MYhref[journalLinks]{http://dx.doi.org/10.1016/j.nuclphysa.2022.122447}{Nuclear
  Physics A
  }\MYhref[journalLinks]{http://dx.doi.org/10.1016/j.nuclphysa.2022.122447}{\textbf{1026}
  (2022) 122447}.

\end{thebibliography}
\end{document}